\documentclass[aps,pra,%
final,%
notitlepage,%
oneside,%
twocolumn,
nobibnotes,%
nofootinbib,%
superscriptaddress,%
centertags,
floatfix,
longbibliography]%
{revtex4-2}
\usepackage{graphicx}
\usepackage{amsmath,amsthm,amssymb,amscd,bm,siunitx}

\DeclareMathOperator{\sinc}{sinc}
\DeclareMathOperator{\rect}{rect}
\usepackage{color}
\usepackage{hyperref}
\newcommand{\pvec}[1]{\vec{#1}\mkern2mu\vphantom{#1}}

\begin{document}

\title{Spatiotemporal entanglement in a noncollinear optical parametric amplifier}

\author{L.~La Volpe}\email{luca.la-volpe@lkb.upmc.fr}
\affiliation{Laboratoire Kastler Brossel, UPMC-Sorbonne Universit\'e, CNRS, ENS-PSL Research University,
Coll\`ege de France, 4 place Jussieu, 75252 Paris, France}
\affiliation{Applied Physics Department, Universit\'e de Gen\`eve, 22 chemin de Pinchat, 1211 Gen\`eve 4, Switzerland}

\author{S.~De}
\affiliation{Laboratoire Kastler Brossel, UPMC-Sorbonne Universit\'e, CNRS, ENS-PSL Research University,
Coll\`ege de France, 4 place Jussieu, 75252 Paris, France}
\affiliation{Applied Physics, Paderborn University, Warburger Strasse 100, 33098 Paderborn, Germany}

\author{M.~I.~Kolobov}
\affiliation{Univ. Lille, CNRS, UMR 8523 - PhLAM - Physique des Lasers Atomes et Mol\'ecules, F-59000 Lille, France}

\author{V.~Parigi}
\affiliation{Laboratoire Kastler Brossel, UPMC-Sorbonne Universit\'e, CNRS, ENS-PSL Research University,
Coll\`ege de France, 4 place Jussieu, 75252 Paris, France}

\author{C.~Fabre}
\affiliation{Laboratoire Kastler Brossel, UPMC-Sorbonne Universit\'e, CNRS, ENS-PSL Research University,
Coll\`ege de France, 4 place Jussieu, 75252 Paris, France}

\author{N.~Treps}
\affiliation{Laboratoire Kastler Brossel, UPMC-Sorbonne Universit\'e, CNRS, ENS-PSL Research University,
Coll\`ege de France, 4 place Jussieu, 75252 Paris, France}

\author{D.~B.~Horoshko}\email{horoshko@ifanbel.bas-net.by}
\affiliation{Univ. Lille, CNRS, UMR 8523 - PhLAM - Physique des Lasers Atomes et Mol\'ecules, F-59000 Lille, France}
\affiliation{B.~I.~Stepanov Institute of Physics, NASB, Nezavisimosti Ave.~68, Minsk 220072 Belarus}%

\date{\today}

\begin{abstract}
We theoretically investigate the generation of two entangled beams of light in the process of single-pass type-I noncollinear frequency degenerate parametric downconversion with an ultrashort pulsed pump. We find the spatio-temporal squeezing eigenmodes and the corresponding squeezing eigenvalues of the generated field both numerically and analytically. The analytical solution is obtained by modeling the joint spectral amplitude of the field by a Gaussian function in curvilinear coordinates. We show that this method is highly efficient and is in a good agreement with the numerical solution. We also reveal that when the total bandwidth of the generated beams is sufficiently high, the modal functions cannot be factored into a spatial and a temporal parts, but exhibit a spatio-temporal coupling, whose strength can be increased by shortening the pump.
\end{abstract}

\maketitle

\section{Introduction}
\label{sec:Intro}

Parametric down-conversion (PDC) of light occurs when a strong coherent pump wave illuminates a nonlinear crystal, where a pump photon at frequency $\omega_p$ is converted into two photons, signal and idler, with frequencies  $\omega_s$ and $\omega_i$ respectively, which sum up to the frequency of the pump photon. When the pump wave is strong enough and a non-degenerate phase-matching condition is satisfied for the three interacting waves, many photon pairs are generated at a time and one obtains a device known as optical parametric amplifier (OPA), capable of amplifying an incoming wave at the signal frequency, which has numerous applications in modern optics. Quantum theory of OPA predicts that even with the vacuum at the input, it generates the signal and idler beams of light which are correlated in photon number \cite{Louisell61} and in field quadratures \cite{Mollow67}. The intensity of the generated beams can be significantly increased by placing the nonlinear crystal inside a cavity resonant at the signal and idler frequencies, a configuration known as optical parametric oscillator (OPO).  Intensity correlations between the two generated ``twin beams'' below the shot noise level were observed first in the OPO configuration \cite{Heidmann87}. Later, it was realized that the correlations of the field quadratures of the twin beams represent an example \cite{Reid88} of the famous Gedankenexperiment of Einstein, Podolsky and Rosen (EPR) \cite{Einstein35}. EPR-correlations of twin beams were observed in PDC light in the OPO \cite{Ou92} and the OPA \cite{Zhang00} configurations, as well as in the four-wave mixing \cite{Boyer08}. It was shown that these correlations represent the fundamental quantum property of entanglement between the two optical modes and, aside from their importance for the foundations of the quantum theory, can be used for quantum teleportation of continuous variables \cite{Braunstein98,Horoshko00}. With the discovery of a possibility to perform a measurement-based quantum computation by making a sequence of measurements on a multimode cluster state, EPR-entangled beams became a key resource for building an optical quantum computer \cite{Asavanant19,Larsen19}.

EPR entanglement of continuous variables is tightly related to quadrature squeezing, which is observed in degenerate OPAs and OPOs. Superimposing two squeezed beams having the same carrier frequency on a balanced beam-splitter, one obtains a pair of EPR-correlated beams at the two beam-splitter outputs \cite{Braunstein98}. This technique is the cornerstone of modern sources of cluster states \cite{Asavanant19,Larsen19}. The inverse process is also possible: combining two EPR-entangled beams having the same carrier frequency on a balanced beam-splitter one obtains two squeezed states at the two outputs \cite{Grangier87}. This technique is useful for characterization of the EPR source \cite{Ou92}, since the measurement of squeezing of one optical beam is often technically simpler than the measurement of EPR correlations of two beams.

\begin{figure}[ht]
\center{\includegraphics[width=\linewidth]{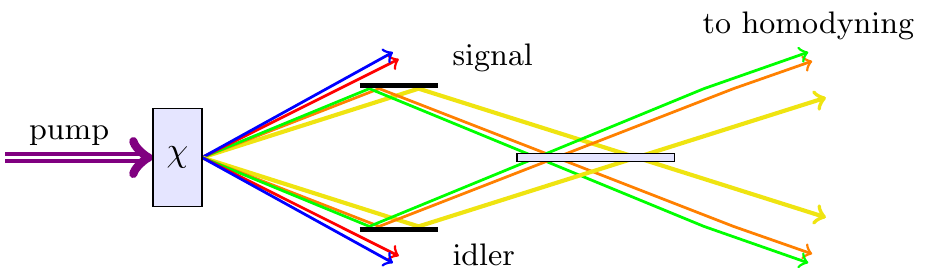}}
\caption{Generation of twin beams via type-I noncollinear PDC. The downconverted field appears in a form of colored cones: each frequency has a distinct angle of perfect phase-matching. A part of this radiation is selected by two symmetrically placed mirrors and recombined on a balanced beam-splitter with a subsequent homodyne detection.}
\label{fig:PDC}
\end{figure}

In this article we explore the possibility of generating two EPR entangled beams in single-pass PDC with a type-I noncollinear phase matching in the sub-picosecond pulsed regime. We are interested in the production of two beams which can be individually addressed and, for instance, manipulated by delay lines and arrays of beam-splitters as necessary for creation of a time-multiplexed cluster state \cite{Asavanant19,Larsen19}. For this purpose the beams need to have the same central frequency and propagate along two distinct directions in space. Since in the type-I PDC the signal and idler fields are generated in a form of colored cones around the direction of propagation of the pump wave, we consider selecting two conjugated directions by two symmetrically placed mirrors, as shown in Fig.~\ref{fig:PDC}.

The simplest way to observe the entanglement of the generated beams is to combine them on a beam splitter with a subsequent homodyne measurement of one of its outputs. Time-domain homodyne measurement of pulsed light is a delicate technique requiring a proper preparation of a local oscillator pulse for precise temporal and spatial mode matching. Measurement of squeezing in the combined beam is the first step towards building a cluster state by time multiplexing \cite{LaVolpe2020}.

In order to obtain a highly multimode configuration, the angular size of the mirrors (seen from the crystal center) is chosen to be bigger than the angular size of the pump. We therefore expect a rich spatial and temporal modal structure of the selected optical beams. One of the main results of our analysis is the demonstration of a strong coupling between the spatial and temporal degrees of freedom of the generated beams, leading to creation of \emph{spatio-temporal modes}, an effect known for photon pairs \cite{Gatti12,Horoshko12} but not yet considered in the high-gain regime. A rich modal structure of entangled beams is highly interesting for adding mode-multiplexing to a time-multiplexed optical cluster state.

An explicit multidimensional multimode analysis of the entangled beams distinguishes our approach from similar works on production of EPR-entangled beams via nonlinear interactions in a pulsed single-pass configuration \cite{Silberhorn01,Wenger05,Shinjo19}. This analysis is based on the decomposition of a Gaussian unitary transformation, like PDC with undepleted pump, into a set of single-mode squeezers for properly defined squeezing eigenmodes \cite{Simon94,Bennink02}, a representation also known as Bloch-Messiah reduction \cite{Braunstein05}. Application of this general procedure to the case of twin beams possesses some remarkable symmetries \cite{Horoshko19}, which simplify the analytic and numerical treatment of an otherwise very complicated six-dimensional problem. On the other hand, our approach is different from the modal decomposition of the full cone of the downconverted light \cite{Migdal10,Perina15}, where the mode-selective detection of squeezing may be very complicated.

The article is structured as follows. In Sec.~\ref{sec:PDC} we consider the unitary transformation of the field in a single-pass PDC with a pulsed pump, applying the full three-dimensional representation of the field. As a result, we find the kernel of the quadratic form, giving the Gaussian transformation generator, and show that it can be made real symmetric in a properly chosen representation picture. In Sec.~\ref{sec:eigenmodes} we show how a Takagi factorization of this kernel allows one to obtain the modal functions of the squeezing eigenmodes in the Fourier domain. We also demonstrate that a Gaussian modeling is possible for this kernel in  curvilinear spatio-temporal coordinates, which allows us to find approximate analytic expressions for the squeezing eigenfunctions and squeezing eigenvalues. We discuss the modal dimensionality of the twin-beams on the basis of the Schmidt number. In Sec.~\ref{sec:sim} we analyze numerically the archetypal example of beta-barium borate (BBO) crystal and find the squeezing eigenfunctions and squeezing eigenvalues, which are remarkably close to that found analytically. Section~\ref{sec:Conclusions} summarizes the results and concludes the article. The appendices contain the details of mathematical calculations.

\section{Noncollinear parametric downconversion}\label{sec:PDC}
\subsection{Notations and the equation of motion}

The model we adopt for the description of single-pass pulsed PDC in a $\chi^{(2)}$ nonlinear crystal is developed in Refs.~\cite{Gatti03,Brambilla04,Caspani10,Horoshko12,Gatti12} on the basis of the wave equation and is recast here in the Hamiltonian form. We consider a crystal slab of length $L$, infinite in the transverse directions, cut for type-I noncollinear phase-matching. We take the $z$ axis as the pump-laser mean propagation direction and indicate with $\vec{x}=(x,y)$ the position coordinates in the transverse plane, where the $y$ axis is taken so that the optical axis of the crystal lies in the $yz$ plane at an angle $\theta_0$ with the $z$ axis.
%
%

The pump is a Gaussian beam focused at position $z_0$ inside the crystal. It is polarized along the $y$ direction and propagates through the crystal as an extraordinary wave. In the time domain it is a Gaussian transform-limited pulse whose maximum passes the position $z_0$ at time $t=0$. Its central frequency is denoted by $\omega_p$. The pump is treated as an undepleted deterministic wave and is described by a c-number function of space and time coordinates. The positive frequency part of the pump field (in photon flux units) can be written as
\begin{eqnarray}\label{eq:FourierPump}
E^{(+)}_p(z,\vec{x},t) =&&\int A_p(\vec{q},\Omega)e^{ik_{pz}(\vec{q},\Omega)(z-z_0) + i\vec{q}\cdot\vec{x}}\\\nonumber
&&\times e^{-i(\omega_p+\Omega)t} \frac{\mathrm{d}\vec{q}}{(2\pi)^2}\frac{\mathrm{d}\Omega}{2\pi},	
\end{eqnarray}
where $\vec{q}=(q_x,q_y)$ is the transverse component of the wave-vector and $\Omega$ represents the frequency offset from the carrier frequency. The pump amplitude $A_p(\vec{q},\Omega)$ does not depend on $z$, since the pump is undepleted. All variations of the pump wave in the longitudinal direction are determined by the longitudinal component of the wave-vector for given $\vec{q}$ and $\Omega$, which is
\begin{equation}\label{eq:kpz}
k_{pz}(\vec{q},\Omega)
=\sqrt{\left(\frac{n_{p}(\vec{q},\omega_p+\Omega)(\omega_p+\Omega)}{c}\right)^2-|\vec{q}|^2},
\end{equation}
where $n_{p}(\vec{q},\omega)$ is the refractive index of the extraordinary wave at frequency $\omega=\omega_p+\Omega$ propagating along the direction determined by the transverse wave-vector $\vec{q}$. It can be expressed via the ordinary and extraordinary refractive indices of a uniaxial crystal (see Appendix~\ref{appendix:nex}).

As a result of nonlinear transformation of the pump field in the crystal, a subharmonic field emerges with the central frequency $\omega_0=\omega_p/2$. In the type-I phasematching, considered here, the subharmonic is polarized in the $xz$ plane and propagates as an ordinary wave. This field is treated in the framework of quantum theory and is described by a Heisenberg operator, being a function of space and time coordinates. The positive frequency part of the Heisenberg field operator (in photon flux units) can be written in a form of Fourier integral:
\begin{eqnarray}\label{eq:FourierSignal}
\hat E^{(+)}(z,\vec{x},t) =&&\int \hat a(z,\vec{q},\Omega)e^{i\vec{q}\cdot\vec{x}-i(\omega_0+\Omega)t} \frac{\mathrm{d}\vec{q}}{(2\pi)^2}\frac{\mathrm{d}\Omega}{2\pi}.			
\end{eqnarray}
Here $\hat a(z,\vec{q},\Omega)$ is the annihilation operator of a photon at position $z$ with the transverse wave-vector $\vec{q}$ and frequency $\omega_0+\Omega$. Evolution of this operator along the crystal is described by the following integro-differential equation \cite{Brambilla04}
\begin{eqnarray}\nonumber
\frac{\partial\hat{a}(z,\vec{q},\Omega)}{\partial z} &&= ik_z(\vec{q},\Omega)\hat{a}(z,\vec{q},\Omega)+ \chi\int A_p(\vec{q}+\vec{q}\,{}',\Omega+\Omega') \\\label{eq:evolution}
&&\times \hat{a}^\dagger(z,\vec{q}\,{}',\Omega')
e^{ik_{pz}(\vec{q},\Omega)(z-z_0)}
\frac{\mathrm{d}\vec{q}\,{}'}{(2\pi)^2}\frac{\mathrm{d}\Omega'}{2\pi},			
\end{eqnarray}
where $\chi$ is the coupling constant and the longitudinal wave-vector of the subharmonic field is
\begin{equation}\label{eq:kz}
k_z(\vec{q},\Omega)=\sqrt{\left(\frac{n_o(\omega_0+\Omega)(\omega_0+\Omega)}{c}\right)^2-|\vec{q}|^2},
\end{equation}
with $n_o(\omega)$ being the refractive index of the ordinary wave at frequency $\omega$.

Equation~(\ref{eq:evolution}) can be rewritten in the Hamiltonian form
\begin{eqnarray}\label{eq:evolutionHam}
\frac{\partial\hat{a}(z,\vec{q},\Omega)}{\partial z} =&&\frac{i}\hbar\left[\hat{a}(z,\vec{q},\Omega),\hat{\mathcal{H}}_0(z)+\hat{\mathcal{V}}(z,z)\right]			\end{eqnarray}
with the linear propagation Hamiltonian
\begin{eqnarray}\label{eq:H0}
\hat{\mathcal{H}}_0(z)&&= \frac{\hbar}{(2\pi)^3}\int k_z(\vec{q},\Omega) \hat{a}^\dagger(z,\vec{q},\Omega) \hat{a}(z,\vec{q},\Omega)
\mathrm{d}\vec{q}\mathrm{d}\Omega,		
\end{eqnarray}
and the parametric interaction Hamiltonian
\begin{eqnarray}\nonumber
&&\hat{\mathcal{V}}(z,z') = \frac{-i\hbar \chi}{2(2\pi)^6}\int A_p(\vec{q}+\vec{q}\,{}',\Omega+\Omega')
e^{ik_{pz}(\vec{q},\Omega)(z-z_0)} \\\label{eq:V}
&&\times \hat{a}^\dagger(z',\vec{q},\Omega) \hat{a}^\dagger(z',\vec{q}\,{}',\Omega')
\mathrm{d}\vec{q}\mathrm{d}\Omega			
\mathrm{d}\vec{q}\,{}'\mathrm{d}\Omega' + H.c.			
\end{eqnarray}
together with the equal-space commutation relations \cite{Huttner90}
\begin{equation}\label{eq:commutator}
\left[\hat{a}(z,\vec{q},\Omega),\hat{a}^\dagger(z,\vec{q}\,{}',\Omega')\right]
= (2\pi)^3\delta(\Omega-\Omega')\delta\left(\vec{q}-\vec{q}\,{}'\right).
\end{equation}

The full Hamiltonian $\hat{\mathcal{H}}_0(z)+\hat{\mathcal{V}}(z,z)$ has a meaning of the $z$-component of the field momentum \cite{Huttner90}. Note that the interaction Hamiltonian (\ref{eq:V}) depends on the position in two ways: directly via the pump dependence and indirectly via the $z$-dependence of the field operator in the Heisenberg picture.

\subsection{Interaction picture \label{subsec:int}}

For a perturbative treatment of the field evolution in the crystal it is convenient to introduce the slowly-varying field amplitude operator $\hat\epsilon(z,\vec{q},\Omega)$ defined as  \cite{Kolobov99}
\begin{eqnarray}\label{eq:epsilon}
\hat a(z,\vec{q},\Omega) = \hat\epsilon(z,\vec{q},\Omega)e^{ik_{z}(\vec{q},\Omega)(z-z_0)}.			\end{eqnarray}
This operator varies slowly with $z$ due to nonlinear coupling of the waves and corresponds to the photon annihilation operator in the interaction picture, introduced by a unitary transformation
\begin{eqnarray}\label{eq:epsilon2}
\hat\epsilon(z,\vec{q},\Omega) = \hat{\mathcal{U}}_I^\dagger(z,0)\hat a(0,\vec{q},\Omega)\hat{\mathcal{U}}_I(z,0),			
\end{eqnarray}
where the interaction picture evolution operator is
\begin{eqnarray}\label{eq:UI}
\hat{\mathcal{U}}_I(z,0) = \hat{\mathcal{U}}_0^\dagger(z-z_0) \hat{\mathcal{U}}(z,0),			
\end{eqnarray}
with $\hat{\mathcal{U}}_0(z) = \exp\left\{\frac{i}{\hbar}\hat{\mathcal{H}}_0(0)z\right\}$ being the operator of linear propagation and $\hat{\mathcal{U}}(z,0)$ being the operator of full evolution in the Heisenberg picture, such that the solution of Eq.~(\ref{eq:evolutionHam}) is  $\hat{a}(z,\vec{q},\Omega)=\hat{\mathcal{U}}^\dagger(z,0) \hat{a}(0,\vec{q},\Omega) \hat{\mathcal{U}}(z,0)$. Peculiarity of the interaction picture introduced by Eq.~(\ref{eq:UI}) consists in the possibility to choose the point $z_0$, where the interaction picture operators coincide with those of the Heisenberg picture, and which we call the ``passage point''. This point does not have to coincide with the beginning of the interaction, $z=0$. As we will see later, one can exploit the symmetries of the setup geometry by placing the passage point $z_0$ at the center of the crystal, where the pump is typically focused.

Substituting Eq.~(\ref{eq:epsilon}) into Eq.~(\ref{eq:evolution}), we obtain an integro-differential equation for the slowly-varying amplitude \cite{Gatti03,Caspani10}
\begin{eqnarray}\label{eq:evolution2}
\frac{\partial\hat\epsilon(z,\vec{q},\Omega)}{\partial z} =&&\chi\int A_p(\vec{q}+\vec{q}\,{}',\Omega+\Omega') \hat\epsilon^\dagger(z,\vec{q}\,{}',\Omega') \\\nonumber
&&\times e^{-i\Delta(\vec{q},\Omega,\vec{q}\,{}',\Omega')(z-z_0)}
\frac{\mathrm{d}\vec{q}\,{}'}{(2\pi)^2}\frac{\mathrm{d}\Omega'}{2\pi},			
\end{eqnarray}
where
\begin{equation}\label{eq:Delta}
\Delta(\vec{q},\Omega,\vec{q}\,{}',\Omega')=k_{z}(\vec{q},\Omega)+k_{z}(\vec{q}\,{}',\Omega') -k_{pz}(\vec{q}+\vec{q}\,{}',\Omega+\Omega')
\end{equation}
is the phase mismatch, determining the efficiency of the downconversion process, in which a pump photon of frequency $\omega_p+\Omega+\Omega'$ with transverse wave-vector $\vec{q}+\vec{q}\,{}'$, splits into two photons of frequencies $\omega_0+\Omega$ and $\omega_0+\Omega'$ with transverse wave-vectors $\vec{q}$ and $\vec{q}\,{}'$ respectively.

Substituting Eq.~(\ref{eq:epsilon2}) into Eq.~(\ref{eq:evolution2}), we obtain an equation for the interaction picture evolution operator $\hat{\mathcal{U}}_I(z,0)$. However, a more compact  equation is obtained for the shifted operator $\hat{\mathcal{U}}_{IS}(z,0)=\hat{\mathcal{U}}_I(z,0) \hat{\mathcal{U}}_0^\dagger(z_0)$:
\begin{equation}\label{eq:equationUIS}
\frac{\mathrm{d}\,\hat{\mathcal{U}}_{IS}(z,0)}{\mathrm{d}z} = \frac{i}{\hbar}\hat{\mathcal{V}}_I(z) \hat{\mathcal{U}}_{IS}(z,0),
\end{equation}
which should be solved with the initial condition $\hat{\mathcal{U}}_{IS}(0,0) = \hat{\mathcal{I}}$, where $\hat{\mathcal{I}}$ is the unity operator. In the above equation the interaction picture interaction Hamiltonian is
\begin{eqnarray}\nonumber
&&\hat{\mathcal{V}}_I(z) = \hat{\mathcal{U}}_0^\dagger(z-z_0)\hat{\mathcal{V}}(z,0)\hat{\mathcal{U}}_0(z-z_0) \\\nonumber
&&= \frac{-i\hbar \chi}{2(2\pi)^6}\int A_p(\vec{q}+\vec{q}\,{}',\Omega+\Omega')
e^{-i\Delta(\vec{q},\Omega,\vec{q}\,{}',\Omega')(z-z_0)} \\\label{eq:VI}
&&\times \hat{a}^\dagger(0,\vec{q},\Omega) \hat{a}^\dagger(0,\vec{q}\,{}',\Omega')
\mathrm{d}\vec{q}\mathrm{d}\Omega			
\mathrm{d}\vec{q}\,{}'\mathrm{d}\Omega' + H.c.			
\end{eqnarray}

The solution of Eq.~(\ref{eq:equationUIS}) describes the field evolution from the crystal input ($z=0$) to the crystal output ($z=L$) and can be written in the form of a $\mathcal{T}$-exponent~\cite{Louisell90}
\begin{align}\label{eq:T-exponent}
	\hat{\mathcal{U}}_{IS}(L,0)=\mathcal{T}e^{\frac{i}{\hbar}\int_0^L dz\, \hat{\mathcal{V}}_I(z)},
\end{align}
where the symbol $\mathcal{T}$ denotes a $z$-ordering operator, putting the operators with higher $z$-values to the left in the expansion of the exponential.

\subsection{Magnus expansion}

Decomposing $\ln\hat{\mathcal{U}}_{IS}(L,0)$ in the Taylor series in the modulus of the coupling constant $|g|$, one can represent the $\mathcal{T}$-exponent in the form of Magnus expansion~\cite{Blanes09,Lipfert18}
\begin{align}\label{eq:Magnus}
	\hat{\mathcal{U}}_{IS}(L,0)=e^{\hat\Xi_1 + \hat\Xi_2 + \hat\Xi_3 + \dots},
\end{align}
where $\hat\Xi_k$ is an operator proportional to $|\chi|^k$, and the first two terms in Eq.~(\ref{eq:Magnus}) are
\begin{eqnarray}\label{Xi1}
	 \hat\Xi_1 &=& \frac{i}\hbar\int_0^L dz\, \hat{\mathcal{V}}_I(z), \\\label{Xi2}
	 \hat\Xi_2 &=& -\frac1{2\hbar^2}\int_0^L dz_1\int_0^{z_1} dz_2\, [\hat{\mathcal{V}}_I(z_1),\hat{\mathcal{V}}_I(z_2)].
\end{eqnarray}

When the coupling of the waves in the nonlinear crystal is not too strong, the Magnus series in the exponent of Eq.~(\ref{eq:Magnus}) converge sufficiently fast. Limiting these series to the first $k$ terms, we obtain the $k$th Magnus approximation for the evolution operator, which we denote by $\hat{\mathcal{U}}_{IS}^{[k]}$. This operator is unitary for any $k$, which is a great advantage of the Magnus expansion as compared to other approximate methods such as the Dyson expansion. It has been shown analytically for continuous-wave PDC \cite{Lipfert18} and numerically for pulsed PDC \cite{Christ13} that the first order of the Magnus expansion is sufficient when the degree of squeezing does not surpass 12 dB. Typical experiments with CV-entangled beams of light are operating at lower squeezing, and thus we limit our consideration to the first-order Magnus approximation, which implies a limitation on the peak pump power. It should be noted that in the noncollinear PDC the wave coupling is relatively weak even for high pump powers, because of fast spatial separation of the subharmonic and the pump beams in the crystal. Thus, the first-order evolution operator $\hat{\mathcal{U}}_{IS}^{[1]}=\exp\hat\Xi_1$ should provide a good description of the field evolution in the crystal.

Substituting Eq.~(\ref{eq:VI}) into Eq.~(\ref{Xi1}), choosing $z_0=L/2$ and performing the integration, we obtain
\begin{eqnarray}\nonumber
\hat\Xi_1 = \frac{\chi L}{2(2\pi)^6}\int &&\mathcal{K}(\vec{q},\Omega,\vec{q}\,{}',\Omega')
\hat{a}^\dagger(0,\vec{q},\Omega) \hat{a}^\dagger(0,\vec{q}\,{}',\Omega')\\\label{eq:Xi1}
&&\times \mathrm{d}\vec{q}\mathrm{d}\Omega			
\mathrm{d}\vec{q}\,{}'\mathrm{d}\Omega' - H.c.,
\end{eqnarray}
where
\begin{eqnarray}\nonumber
\mathcal{K}(\vec{q},\Omega,\vec{q}\,{}',\Omega') &=& A_p(\vec{q}+\vec{q}\,{}',\Omega+\Omega')\\\label{eq:Kernel}
&\times& \sinc\left(\Delta(\vec{q},\Omega,\vec{q}\,{}',\Omega')L/2\right)
\end{eqnarray}
is the squeezing kernel, corresponding to the squeezing matrix in the case of discrete modes \cite{Bennink02}. This rather simple form of the squeezing kernel is a result of our choice of the passage point to the interaction picture $z_0$. A further simplification is related to the form of the term $A_p(\vec{q},\Omega)$, which is the transverse and temporal Fourier transform of the pump field envelope at $z_0$. If the pump beam is focused at the crystal center and is transform-limited at this point, then the function $A_p(\vec{q},\Omega)$ is real (or has a constant phase). Assuming Gaussian shapes for the spatial and temporal distribution of the pump beam we can write the pump amplitude as
\begin{equation}\label{eq:pumpamp}
A_p(\vec{q},\Omega) = A_0 \exp\left(-\frac{|\vec{q}|^2}{4 q_p^2}-\frac{\Omega^2}{4\Omega_p^2}\right)
\end{equation}
where $A_0$ is the peak pump amplitude (in the photon flux units), while $q_p$ and $\Omega_p$ are (intensity) standard deviations of the spatial spectrum and the frequency spectrum respectively. The phases of $A_0$ and $\chi$ can be removed by a trivial phase shift $\hat{a}(0,\vec{q},\Omega)\rightarrow \hat{a}(0,\vec{q},\Omega)\exp[i\arg(\chi A_0)/2]$, and below, without loss of generality, we will take both these parameters real and positive.

The squeezing kernel, Eq.~(\ref{eq:Kernel}), is real in this case, which greatly simplifies the modal analysis of the generated light.

\section{Squeezing eigenmodes \label{sec:eigenmodes}}
\subsection{Takagi factorization}

The interaction picture evolution operator $\exp\hat\Xi_1$ with its generator defined by Eq.~(\ref{eq:Xi1}) represents a multimode squeezing operator \cite{Bennink02}. It means that the subharmonic field at the output of the nonlinear crystal is in a multimode squeezed state. By a proper choice of the modal basis for this field the output state can be represented as direct product of single-mode squeezed states for each spatio-temporal mode. Such a representation can be based on a diagonalization of the covariance matrix \cite{Shapiro97}, on a Takagi factorization of the squeezing matrix \cite{Bennink02} or on the more general formalism of the Bloch-Messiah reduction of a Gaussian unitary transformation \cite{Arvind95,Braunstein05,Wasilewski06,Cariolaro16}. The equivalence of all these approaches has been shown in Ref.~\cite{Horoshko19}. Following Bennink and Boyd \cite{Bennink02} we introduce the squeezing eigenmodes by Takagi factorization of the squeezing kernel, Eq.~(\ref{eq:Kernel}). Note that this kernel is symmetric with respect to variable exchange $\{\vec{q},\Omega\}\leftrightarrow\{\vec{q}\,{}',\Omega'\}$, which follows from the signal-idler symmetry in a frequency-degenerate type-I PDC. As a consequence, its singular value decomposition can be written using the Takagi factorization:
\begin{equation}\label{eq:Takagi}
\mathcal{K}(\vec{q},\Omega,\vec{q}\,{}',\Omega') = \sum_m\sigma_m f_m(\vec{q},\Omega)f_m(\vec{q}\,{}',\Omega'),
\end{equation}
where the non-negative numbers $\sigma_m$ are the singular values of the kernel and the complex functions $f_m(\vec{q},\Omega)$ create a complete orthonormal set of functions in the space of functions of $\{\vec{q},\Omega\}$. The above factorization is always possible for a symmetric square-integrable kernel \cite{HornJohnson}. To guarantee the square integrability of the kernel below we accept that the subharmonic field is spatially filtered, so that only a limited range of values $\vec{q}$ is taken into consideration. For a real kernel the functions $f_m(\vec{q},\Omega)$ possess a remarkable property: they are either real of purely imaginary \cite{Horoshko19}.

Defining the new modes by the modal functions $f_m(\vec{q},\Omega)$ and the corresponding annihilation operators
\begin{equation}\label{eq:b}
\hat b_m = \frac{1}{(2\pi)^3}\int f_m^*(\vec{q},\Omega)\hat a(0,\vec{q},\Omega)\mathrm{d}\vec{q}\mathrm{d}\Omega,
\end{equation}
we rewrite the interaction picture evolution operator as
\begin{eqnarray}\label{eq:modal}
e^{\hat\Xi_1} = e^{\frac12 \chi L\sum_m\sigma_m\left(\hat{b}_m^{\dagger2}-\hat{b}_m^2\right)},
\end{eqnarray}
which is a product of squeezing operators for every mode. The degree of squeezing of the $m$th mode is $r_m=\chi L\sigma_m$.

The operator of the full evolution in the Heisenberg picture reads
\begin{eqnarray}\label{eq:modal2}
\hat{\mathcal{U}}(L,0) &=& \hat{\mathcal{U}}_0(L/2) \hat{\mathcal{U}}_{IS}(L,0) \hat{\mathcal{U}}_0(L/2)\\\nonumber
&=&e^{\frac{i}{\hbar}\hat{\mathcal{H}}_0(0)L/2} e^{\hat\Xi_1} e^{\frac{i}{\hbar}\hat{\mathcal{H}}_0(0)L/2}\\\nonumber
&=&e^{\frac12 \chi L\sum_m\sigma_m\left(\hat{c}_m^{\dagger2}-\hat{c}_m^2\right)} e^{\frac{i}{\hbar}\hat{\mathcal{H}}_0(0)L},
\end{eqnarray}
where the operators $\hat c_m$ correspond to modes with the modal functions $\bar f_m(\vec{q},\Omega)=f_m(\vec{q},\Omega)\exp(ik_z(\vec{q},\Omega)L/2)$.

When the operator (\ref{eq:modal2}) acts on the vacuum state of the field, the free evolution operator $e^{\frac{i}{\hbar}\hat{\mathcal{H}}_0(0)L}$ leaves the vacuum unaffected. Thus, the output field of an unseeded parametric amplifier consists of squeezed modes defined by the modal functions $\bar f_m(\vec{q},\Omega)$, which differ from the functions $f_m(\vec{q},\Omega)$ by a phase factor, corresponding to dispersive propagation from the center of the crystal to its output plane. The output modal functions $\bar f_m(\vec{q},\Omega)$ are complex functions and are not easy to visualize. In the following we prefer to work with the modal functions $f_m(\vec{q},\Omega)$ which are either purely real or purely imaginary and which correspond to the squeezing eigenmodes of the parametric amplifier backward projected to the crystal center.

\subsection{Reduction of the Takagi factorization to the spectral decomposition \label{sec:reduction}}

Since the kernel $\mathcal{K}(\vec{q},\Omega,\vec{q}\,{}',\Omega')$ is real and symmetric it can be diagonalized in its eigenbasis:
\begin{equation}\label{eq:orthogonal}
\mathcal{K}(\vec{q},\Omega,\vec{q}\,{}',\Omega') = \sum_m\lambda_m w_m(\vec{q},\Omega)w_m(\vec{q}\,{}',\Omega'),
\end{equation}
where $\lambda_m$ are real eigenvalues and $w_m(\vec{q},\Omega)$ are real eigenfunctions of the kernel. The difference with the Takagi factorisation, Eq.~(\ref{eq:Takagi}), is twofold: first, the eigenvalues $\lambda_m$ can be negative, while the singular values $\sigma_m$ are always non-negative, and second, the eigenfunctions $w_m(\vec{q},\Omega)$ are real, while the functions $f_m(\vec{q},\Omega)$ are either real or imaginary. However, the connection between these two decompositions is simple \cite{Horoshko19}. For any $m$ such that $\lambda_m$ is nonengative, $\sigma_m=\lambda_m$ and $f_m(\vec{q},\Omega)=w_m(\vec{q},\Omega)$. For any $m$ such that $\lambda_m$ is negative, $\sigma_m=-\lambda_m$ and $f_m(\vec{q},\Omega)=iw_m(\vec{q},\Omega)$.

Real spectral decomposition will be our principal tool in the numerical calculation of the squeezing eigenmodes in the subsequent sections.

\subsection{Approximations for the phase matching \label{subsec:approximations}}

At certain regimes of PDC, the numerically found squeezing eigenmodes can be compared to those found by Gaussian modeling of the squeezing kernel and a subsequent Takagi factorisation, Eq.~(\ref{eq:Takagi}), in an analytic form. Gaussian modeling of the kernel requires application of the nearly plain-wave and monochromatic pump approximation (NPMPA) \cite{Caspani10,Horoshko12,Gatti12}.

To formulate this approximation let us introduce new variables \cite{Caspani10}: $\vec{q}_+=\vec{q}+\vec{q}\,{}'$, $\vec{q}_-=\frac12\left(\vec{q}-\vec{q}\,{}'\right)$, $\Omega_+=\Omega+\Omega'$, $\Omega_-=\frac12\left(\Omega-\Omega'\right)$. In these variables the squeezing kernel, Eq.~(\ref{eq:Kernel}), with the pump defined by Eq.~(\ref{eq:pumpamp}) reads
\begin{eqnarray}\label{eq:Kernel+}
&&\mathcal{K}(\vec{q},\Omega,\vec{q}\,{}',\Omega') = A_0 \exp\left(-\frac{|\vec{q}_+|^2}{4 q_p^2}-\frac{\Omega_+^2}{4\Omega_p^2}\right)\\\nonumber
&&\times \sinc\left[\Delta\left(\frac{\vec{q}_+}2+\vec{q}_-, \frac{\Omega_+}2+\Omega_-, \frac{\vec{q}_+}2-\vec{q}_-, \frac{\Omega_+}2-\Omega_-\right)\frac{L}2\right].
\end{eqnarray}
Due to the exponential factor this function is  non-zero only in a small region around $(\vec{q}_+,\Omega_+)=0$. The size of this region is determined by the standard deviations $q_p$ and $\Omega_p$ of the pump. For sufficiently wide and long pump pulse these deviations are so small that the sinc factor in Eq.~(\ref{eq:Kernel+}) can be considered as constant in the variables $(\vec{q}_+,\Omega_+)$ and evaluated at the point where they are zero, which means application of NPMPA. A rigorous formulation of the conditions for this approximation in terms of dispersive properties of the crystal can be found in Ref.~\cite{Caspani10}.

\begin{figure}[htbp] 
\center{\includegraphics[width=\linewidth]{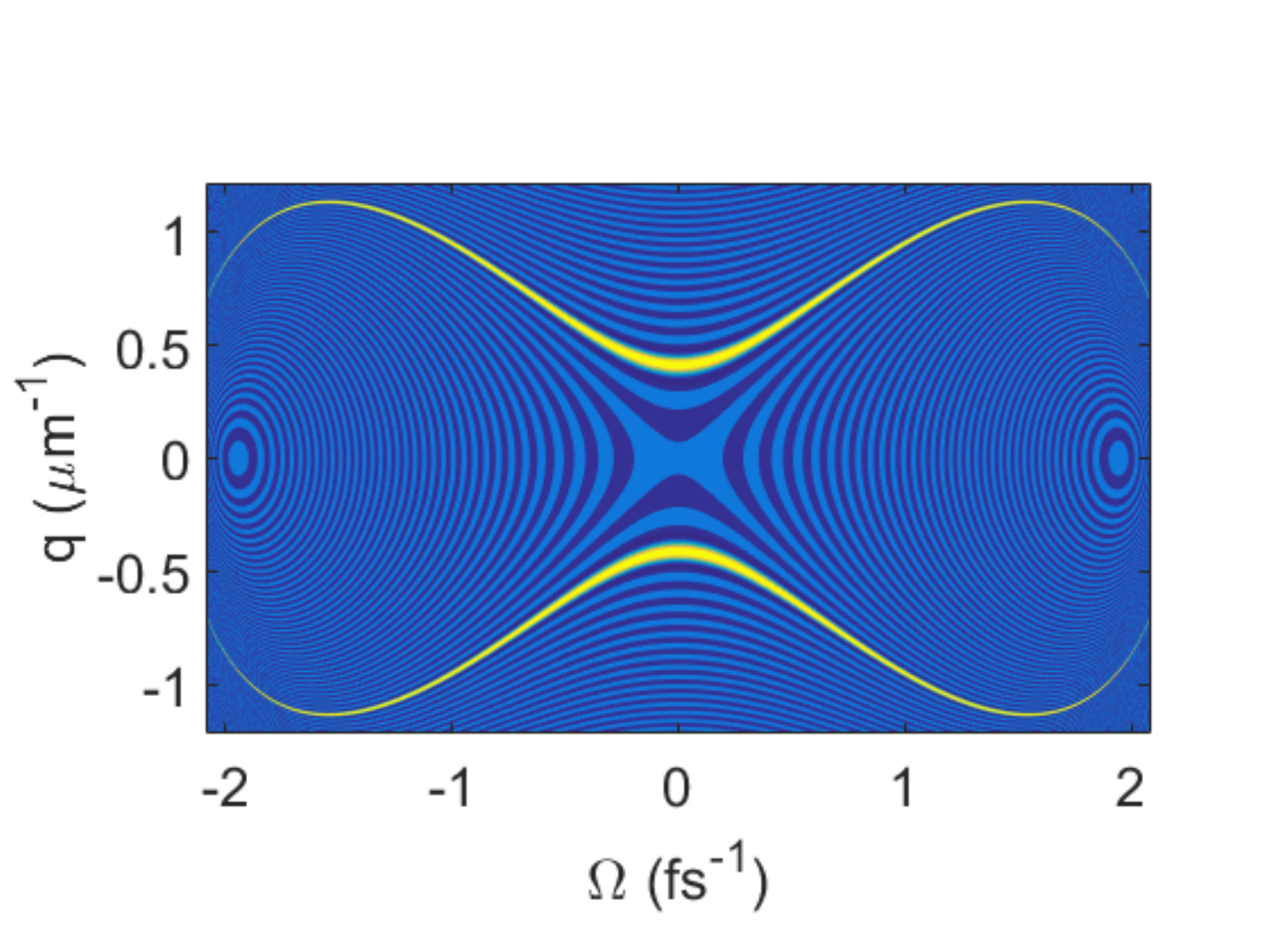}}
\caption{The phase-matching function for the BBO crystal cut for frequency-degenerate noncollinear type-I PDC at the angle $\theta_0=\ang{29.62}$ pumped at the wavelength 397.5 nm.}
\label{fig:phasematch-noncoll}
\end{figure}

In NPMPA the squeezing kernel takes a simple form
\begin{eqnarray}\nonumber
\mathcal{K}_0(\vec{q},\Omega,\vec{q}\,{}',\Omega') &&= A_0 \exp\left(-\frac{|\vec{q}_+|^2}{4 q_p^2}-\frac{\Omega_+^2}{4\Omega_p^2}\right) \Phi_0(\vec{q}_-,\Omega_-),\\\label{eq:K0}
\end{eqnarray}
where
\begin{eqnarray}\label{eq:Phi0}
\Phi_0(\vec{q},\Omega) = \sinc\left[\Delta\left(\vec{q},\Omega, -\vec{q},-\Omega\right)L/2\right]
\end{eqnarray}
is the phase-matching function. For the considered case of type-I PDC this function is independent of the direction of the vector $\vec q$ and depends only on its modulus $q=|\vec q|$, as can be easily seen from Eqs.~(\ref{eq:kz}), (\ref{eq:Delta}), and (\ref{eq:Phi0}). The shape of the phase-matching function in the plane $(q,\Omega)$ is calculated for a beta-barium borate (BBO) crystal with the help of the Sellmeier equation for dispersion and is shown in Fig.~\ref{fig:phasematch-noncoll}. This function takes its maximal value 1 along the line in the $(q,\Omega)$ plane (phase-matched curve) where the phase mismatch $\Delta\left(\vec{q},\Omega, -\vec{q},-\Omega\right)$ is zero.

The phase-matching function can be expressed in elementary functions by making the paraxial and quadratic dispersion approximation (PQDA), which consists in keeping only terms up to the second order in the Taylor expansion of the $z$-component of the ordinary wave-vector as function of $\vec{q}$ and $\Omega$:
\begin{equation}\label{eq:kz-Taylor}
\begin{aligned}
k_z(\vec{q},\Omega)&=\sqrt{k(\Omega)^2-q_x^2-q_y^2}\\
&\approx k_0+k_0'\Omega + \frac12 k_0''\Omega^2 - \frac1{2k_0}\left(q_x^2+q_y^2\right),
\end{aligned}
\end{equation}
where $k(\Omega)=n_o(\omega_0+\Omega)(\omega_0+\Omega)/c$ is the modulus of the ordinary wave-vector at frequency $\omega_0+\Omega$, while $k_0$, $k_0'$ and $k_0''$ are its value and two derivatives at $(\vec{q},\Omega)=0$. PQDA is valid if the spatial filtering limits the maximal value of $|\pvec{q}|$, and as a consequence, the maximal value of $|\Omega|$, which is related to $|\pvec{q}|$ via the phase-matching function.

Now, from Eqs.~(\ref{eq:Delta}) and (\ref{eq:kz-Taylor}) we obtain the phase mismatch in PQDA and NPMPA
\begin{eqnarray}\nonumber
\Delta&(\vec{q},\Omega,-\vec{q},-\Omega)=k_{z}(\vec{q},\Omega)+k_{z}(-\vec{q},-\Omega) -k_{pz}(0,0)\\\label{eq:Delta2}
&\approx 2k_0-k_p + k_0''\Omega^2 - \frac1{k_0}\left(q_x^2+q_y^2\right),
\end{eqnarray}
where $k_p=k_{pz}(0,0)$ is the wave-vector of the pump at frequency $\omega_p$ directed along the $z$ axis. Introducing the characteristic spectral and spatial widths of the phase-matching function \cite{Gatti03,Brambilla04,Caspani10,Horoshko12,Gatti12} $\Omega_0=\sqrt{1/(k_0''L)}$ and $Q_0=\sqrt{k_0/L}$ and the collinear mismatch phase $\gamma=(2k_0-k_p)L/2$, we write the phase-matching function in the following PQDA form:
\begin{eqnarray}\label{eq:Phi02}
\Phi_0(\vec{q},\Omega) \approx \sinc\left(\gamma+\frac{\Omega^2}{2\Omega_0^2}-\frac{q_x^2+q_y^2}{2Q_0^2}\right).
\end{eqnarray}

Noncollinear configuration corresponds to positive $\gamma$. In this case, the degenerate frequency $\Omega=0$ is perfectly matched on the cone defined by the relation $|\pvec{q}|=q_d$, where $q_d=\sqrt{2\gamma}Q_0$. This cone is characterized by the angle $\theta_s$ between any of its directions and the $z$ axis, such that $q_d=k_0\sin\theta_s\approx k_0\theta_s$. The exact condition of zero phase mismatch is $k_p=2k_0\cos\theta_s$. As a consequence, in PQDA one can write $\gamma=k_0L(1-\cos\theta_s)\approx k_0L\theta_s^2/2$.

For every $\vec{q}$ such that $|\pvec{q}|\ge q_d$, there are two perfectly matched frequencies $\pm\Omega^{(pm)}(\vec{q})$, where
\begin{equation}\label{eq:Opm}
\Omega^{(pm)}(\vec{q}) = \Omega_0\sqrt{\frac{q_x^2+q_y^2}{Q_0^2}-2\gamma}.
\end{equation}
This expression shows that in the framework of PQDA the perfect phase-matched surface in the $(\vec{q},\Omega)$ space has the shape of a hyperboloid. The cross-section of this hyperboloid at $q_y=0$ gives the phase-matched (yellow) areas in Fig.~\ref{fig:phasematch-noncoll}. At high values of $|\Omega|$ the perfect phase-matched curve deflects from the shape of hyperbola because PQDA is not valid any more and higher orders of the Taylor expansion in Eq.~(\ref{eq:kz-Taylor}) become important.

Note that the applicability of PQDA is determined by the filtering of the downconverted light, while that of NPMPA is determined by the pump pulse size. In Appendix~\ref{appendix:npmpa} we analyze the conditions for applicability of NPMPA when PQDA is valid.

\subsection{Joint spectral amplitude}
\label{sec:JSA}
\begin{figure}[ht!] 
\center{\includegraphics[width=1\linewidth]{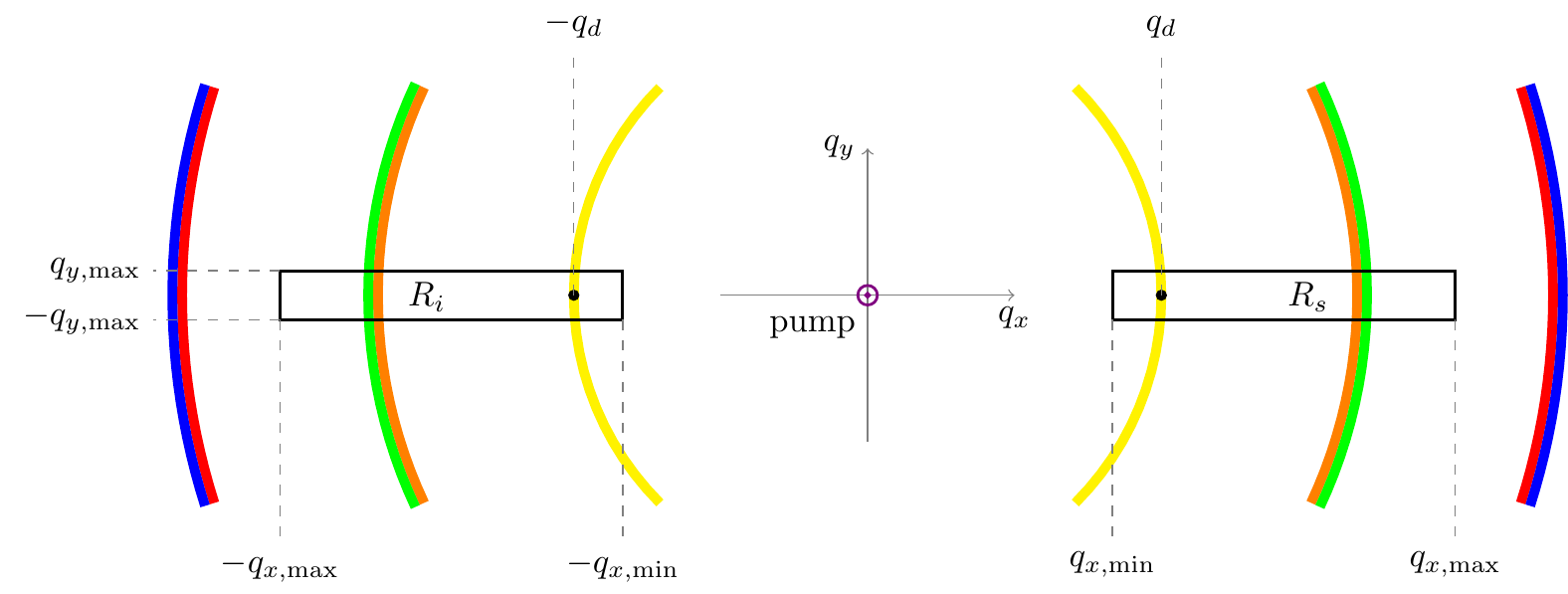}}
\caption{Schematic representation of the signal region $R_s$ and the idler one $R_i$ selected by two mirrors in the OPA output. The colored arcs show spectral components emitted at given angle and correspond to the colored arrows in Fig.~\ref{fig:PDC}. $q_d$ is the transverse wave vector corresponding to perfect phase-matching at the degenerate frequency $\omega_0=\omega_p/2$.}
\label{fig:Mirrors}
\end{figure}

As we have seen from the analysis of the preceding section, in a type-I noncollinear PDC the subharmonic radiation emerges in a form of ``colored cones''. In practice, only a part of this radiation may be interesting. As explained in the Introduction, in this study we consider the case where two mirrors select two regions in the $\vec{q}$ space for subsequent optical processing. One region, $R_s$, corresponds to $q_x\in [q_{x,\mathrm{min}},q_{x,\mathrm{max}}]$ and $q_y\in [-q_{y,\mathrm{max}},q_{y,\mathrm{max}}]$, and the field emitted in this direction is referred to as ``signal'', see Fig.~\ref{fig:Mirrors}.

The other region, $R_i$, corresponds to $q_x\in [-q_{x,\mathrm{max}},-q_{x,\mathrm{min}}]$ and $q_y\in [-q_{y,\mathrm{max}},q_{y,\mathrm{max}}]$, and the field emitted in this direction is referred to as ``idler''. The limits of the regions are chosen to comprise maximally the phase-matched area. The idler radiation has the transverse wave-vector opposite to that of the signal one, and in this sense the signal and the idler areas are conjugated: every photon in the signal field has in the idler field its ``twin'', emerged in the same elementary act of photon-pair creation.

Mathematically, the signal region $R_s$ is described by its indicator function
\begin{equation}\label{eq:Pi}
\Pi(\vec{q}) =\left\{\begin{array}{cl} 1, & \text{if} \,\,\,\vec{q}\in R_s,\\
                                       0, & \text{otherwise}, \end{array}\right.
\end{equation}
while the idler region $R_i$ has the indicator function $\Pi(-\pvec{q})$. The squeezing kernel of the light reflected by the mirrors is given by Eq.~(\ref{eq:Kernel}) multiplied by the indicator functions of the signal and idler regions:
\begin{eqnarray}\label{eq:KJ}
\mathcal{K}(\vec{q},\Omega,\vec{q}\,{}',\Omega') &=& \mathcal{J}(\vec{q},\Omega,\vec{q}\,{}',\Omega') + \mathcal{J}(\vec{q}\,{}',\Omega',\vec{q},\Omega),
\end{eqnarray}
where
\begin{equation}\label{eq:J}
\mathcal{J}(\vec{q},\Omega,\vec{q}\,{}',\Omega') = \mathcal{K}(\vec{q},\Omega,\vec{q}\,{}',\Omega')\Pi(\pvec{q})\Pi(-\vec{q}\,{}')\Pi_0(\Omega)\Pi_0(\Omega')
\end{equation}
is a kernel localized in the region $\vec{q}\in R_s$, $\vec{q}\,{}'\in R_i$. Here $\Pi_0(\Omega)=\rect(|\Omega|/2\Omega_\mathrm{max})$ is a function cutting off the frequencies above the maximal on $R_s$ phase-matched frequency $\Omega_\mathrm{max}=\Omega^{(pm)}(q_{x,\mathrm{max}},0)$. This filtering function is redundant in Eq.~(\ref{eq:J}), since above $\Omega_\mathrm{max}$ the squeezing kernel is almost zero on $R_s$. However, in the next section we will omit the filtering in the $q_x$ direction and the filtering in frequency will be necessary for square-integrability of the kernel. In other words, we replace the spatial filtering by the frequency filtering, which is possible because every frequency component is emitted at a given angle. Filtering in frequency allows us also to define the 3-dimensional regions for the signal field $\bar R_s = \{\vec{q}\in R_s, |\Omega|\le\Omega_\mathrm{max}\}$ and for the idler one $\bar R_i = \{\vec{q}\in R_i, |\Omega|\le\Omega_\mathrm{max}\}$.

The kernel $\mathcal{J}(\vec{q},\Omega,\vec{q}\,{}',\Omega')$ gives the coupling strength of the signal photon with values $(\vec{q},\Omega)$ to the idler photon with values $(\vec{q}\,',\Omega')$ and is generally known in the literature on the low-gain regime of PDC as joint spectral amplitude (JSA) of the photon pair \cite{Law00}. In the high-gain regime, considered here, many photons are emitted at a time, but in the first order of the Magnus expansion their coupling is determined by the same JSA, as in the low-gain regime.

It is known \cite{Horoshko19} that in the case of twin beams the modal functions of squeezing eigenmodes, similar to the squeezing kernel, are localized in the regions $\bar R_s$ and $\bar R_i$. In each of these regions they are proportional to the modal functions of the corresponding Schmidt modes \cite{Law00}, obtained from a singular value decomposition of the JSA:
\begin{equation}\label{eq:SVD}
\mathcal{J}(\vec{q},\Omega,\vec{q}\,{}',\Omega') = \sum_\ell s_\ell c_\ell(\vec{q},\Omega)d_\ell^*(\vec{q}\,{}',\Omega'),
\end{equation}
where $s_\ell$ are the singular values and $c_\ell(\vec{q},\Omega)$ and $d_\ell(\vec{q}\,{}',\Omega')$ are the signal and idler singular eigenfunctions respectively, which are defined up to global phase and can be made both real or purely imaginary for given $\ell$ due to the symmetries of the kernel, as discussed in Sec.~\ref{sec:reduction}.

The modal functions of the squeezing eigenmodes read \cite{Horoshko19}
\begin{align}\label{eq:f1}
f_{\ell}^+(\vec{q},\Omega) &= \frac1{\sqrt{2}}\left[
c_\ell(\vec{q},\Omega) +           d_\ell^*(\vec{q},\Omega)\right], \\\nonumber
f_{\ell}^-(\vec{q},\Omega) &= \frac{i}{\sqrt{2}}\left[ c_\ell(\vec{q},\Omega)
   -d_\ell^*(\vec{q},\Omega)\right]
\end{align}
and are also real or purely imaginary. Both these eigenmodes correspond to the same squeezing eigenvalue $\sigma_{\ell}=s_\ell$, having a multiplicity of 2. The squeezing mode index $m$ in Eq.~(\ref{eq:Takagi}) can be understood as composed from the singular mode index $\ell$ and a binary index corresponding to the choice of the sign $\pm$. When all singular values $s_\ell$ are different, the squeezing eigenmodes defined by Eq.~(\ref{eq:f1}) are unique up to sign.

The singular value decomposition of the JSA is preferable to the Takagi factorization of the squeezing kernel for two reasons. First, since the regions $\bar R_s$ and $\bar R_i$ are well separated in the $(\vec{q},\Omega)$ space, the number of points in the discretized squeezing kernel can surpass by several orders (in each dimension) the number of points necessary to resolve the variations of this kernel in the regions where it is essentially non-zero. The six-dimensional squeezing kernel may thus become highly sparse and untreatable numerically without invoking special methods of sparse array treatment. Second, in many cases the JSA can be modeled by a set of double-Gaussians in each pair of dimensions, and its singular value decomposition can be analytically found by applying the Mehler's formula, as shown in the next section.

\subsection{Gaussian model for JSA in curvilinear coordinates \label{Gaussian}}

To obtain a Gaussian model for the JSA defined by Eq.~(\ref{eq:J}), we note that when the separation of the signal and the idler regions is much larger than the size of these regions, the term $q_x^2/(2Q_0^2)$ is dominant in the argument of sinc in Eq.~(\ref{eq:Phi02}). Therefore, with a good degree of approximation, we can write the equation of the perfectly phase-matched surface in $\bar R_s$ by expressing $q_x$ through the other two variables and leaving only terms up to the lowest (second) order in $\Omega/\Omega_0$ and $q_y/Q_0$: $q_x\approx q_x^{(pm)}(q_y,\Omega)$, where
\begin{equation}\label{eq:Qpm}
q_x^{(pm)}(q_y,\Omega) = q_d + \frac{Q_0^2}{2\Omega_0^2q_d}\Omega^2 - \frac{1}{2q_d}q_y^2.
\end{equation}
This approximation means a replacement of a segment of hyperboloid by a segment of paraboloid. Note that the perfect phase-matched surface in the idler region $\bar R_i$ is $q_x'\approx -q_x^{(pm)}(q_y',\Omega')$. Now we change the coordinates to the curvilinear ones $(q_x,q_y,\Omega)\to(\eta,q_y,\Omega)$, where
\begin{equation}\label{eq:eta}
\eta = \left\{\begin{array}{ll}
q_x - q_x^{(pm)}(q_y,\Omega),& \textrm{if} \,\,q_x\ge0,\\
q_x + q_x^{(pm)}(q_y,\Omega),& \textrm{if} \,\,q_x<0.
\end{array}\right.
\end{equation}
In these coordinates the perfectly phase-matched surfaces in $\bar R_s$ and $\bar R_i$ are defined by the equation $\eta=0$ and the phase-matching function, Eq.~(\ref{eq:Phi02}) can be written as
\begin{eqnarray}\label{Phi03}
\Phi_0(\vec q_-,\Omega_-) \approx \sinc\left(\frac{(\eta-\eta')q_d}{2Q_0^2}\right),
\end{eqnarray}
where we have disregarded the second-order terms in all coordinates compared to the first-order term.  Besides, we make a replacement $\sinc(x)\approx e^{-x^2/(2\sigma_s^2)}$, where $\sigma_s=1.61$ is chosen so that these two functions have the same width at half-maximum \cite{Grice01}. Finally, we replace each rectangle filtering function by a Gaussian, whose $\sigma$-region coincides with the rectangle width multiplied by $\mu$, a fitting parameter allowing us to correctly describe the filtering process. All these approximations allow us to write the JSA defined by Eqs.~(\ref{eq:K0}), (\ref{eq:J}), and (\ref{Phi03}) in the curvilinear coordinates as
\begin{eqnarray}\nonumber
\tilde{\mathcal{J}}_0&&(\eta,q_y,\Omega,\eta',q_y',\Omega') = A_0 \exp\left(-\frac{(\Omega+\Omega')^2}{4\Omega_p^2}\right)\\\nonumber
&&\times\exp\left(-\frac{(\eta+\eta')^2+(q_y+q_y')^2}{4q_p^2}\right)
\exp\left(-\frac{(\eta-\eta')^2}{4\eta_s^2}\right)\\\label{eq:J0}
&&\times\exp\left(-\frac{(q_y-q_y')^2}{4\mu^2q_{y,\mathrm{max}}^2} -\frac{(\Omega-\Omega')^2}{4\mu^2\Omega_\mathrm{max}^2}\right),
\end{eqnarray}
where the tilde denotes a function written in the curvilinear coordinates, $\eta_s=\sqrt{2}\sigma_sQ_0^2/q_d$ is the (intensity) standard deviation in the $\eta-\eta'$ direction, and we have replaced the filtering in the $\eta$ and $\eta'$ directions by filtering in $\Omega$ and $\Omega'$, as discussed in  Sec.~\ref{sec:JSA}. Deriving the filtering terms of Eq.~(\ref{eq:J0}) we have taken into account that the filtering bands in the $q_y+q_y'$ and $\Omega+\Omega'$ directions are much wider than the corresponding pump bandwidths and can be put to infinity.

The obtained JSA, Eq.~(\ref{eq:J0}), is a product of three double-Gaussian functions. The singular value decomposition of a double-Gaussian function is given by the Mehler's formula \cite{Grice01,Wasilewski06,Lvovsky07,Patera10,Horoshko12,Horoshko19}. We show in Appendix~\ref{appendix:Mehler}, that applying this formula to Eq.~(\ref{eq:J0}) and passing back to the Cartesian coordinates, we obtain a decomposition of the JSA defined by Eq.~(\ref{eq:J}) in the form of Eq.~(\ref{eq:SVD}), with a composite index $\ell=(i,j,k)$ and the singular values
\begin{equation}\label{eq:sl}
s_\ell = \mathcal{N} \xi_x^i \xi_y^j \xi_t^k,
\end{equation}
where
\begin{eqnarray}\nonumber
\xi_x&=&(\eta_s-q_p)/(\eta_s+q_p),\\\label{xi}
\xi_y&=&(\mu q_{y,\mathrm{max}}-q_p)/(\mu q_{y,\mathrm{max}}+q_p),\\\nonumber
\xi_t&=&(\mu\Omega_\mathrm{max}-\Omega_p)/(\mu\Omega_\mathrm{max}+\Omega_p),\\\nonumber
\mathcal{N}&=&\pi^\frac32 A_0 \sqrt{(1-\xi_x^2)(1-\xi_y^2)(1-\xi_t^2)}/(uv\tau).
\end{eqnarray}
The $\xi$-parameter for each dimension can be written in the general form $\xi_a=(r_a-1)/(r_a+1)$, where $a$ takes values $x,y,t$ and $r_a$ is the ratio of the larger standard deviation to the smaller one for the corresponding dimension in Eq.~(\ref{eq:J0}), i.e. $r_x=\eta_s/q_p$, $r_y=\mu q_{y,\mathrm{max}}/q_p$, $r_t=\mu\Omega_\mathrm{max}/\Omega_p$. The modal functions of the signal and idler Schmidt modes are respectively
\begin{eqnarray}\nonumber
c_\ell(q_x,q_y,\Omega) &=& h_i\left(uq_x - uq_d - g\tau^2\Omega^2 + \frac{u}{2q_d}q_y^2\right)\\\label{eq:cl2}
&\times& h_j(vq_y)h_k(\tau\Omega) \sqrt{uv\tau},\\\nonumber
d_\ell(q_x,q_y,\Omega) &=& h_i\left(uq_x + uq_d + g\tau^2\Omega^2 - \frac{u}{2q_d}q_y^2\right)\\\label{eq:dl2}
&\times& h_j(vq_y)h_k(\tau\Omega) (-1)^{i+j+k}\sqrt{uv\tau},
\end{eqnarray}
where $g$ is the dimensionless spatio-temporal coupling constant defined as
\begin{equation}\label{eq:g}
g = \frac{uQ_0^2}{2\tau^2\Omega_0^2q_d},
\end{equation}
while $h_n(x) = \left(2^nn!\sqrt{\pi}\right)^{-\frac12}H_n(x)e^{-x^2/2}$ is the Hermite-Gauss function, $H_n(x)$ being the Hermite polynomial. The parameters $u=(\eta_sq_p)^{-\frac12}$, $v=(\mu q_{y,\mathrm{max}}q_p)^{-\frac12}$ and $\tau=(\mu\Omega_\mathrm{max}\Omega_p)^{-\frac12}$ are characteristic widths of the Schmidt modes in the transverse coordinates and the time respectively.

The signal modal functions, Eq.~(\ref{eq:cl2}), are orthonormal and complete on the three-dimensional Hilbert space (see Appendix~\ref{appendix:ortho}) and exhibit correlations between the spatial and temporal degrees of freedom. The same concerns the idler modal functions, Eq.~(\ref{eq:dl2}). These functions can be combined to obtain the modal functions of the squeezing eigenmodes, as shown by Eqs.~(\ref{eq:f1}). In Sec.~\ref{sec:sim} below they are compared to the modal functions obtained by a numerical decomposition of the squeezing kernel, Eq.~(\ref{eq:orthogonal}).

\subsection{Multidimensional Schmidt analysis \label{subsec:Schmidt}}
The analytic expression for the singular eigenvalues, Eq.~(\ref{eq:sl}), allows us to define the effective number of modes in each of the two entangled beams. The commonly used approach consists in calculating the Schmidt number \cite{Law04}:
\begin{equation}\label{eq:Ks}
K=\frac{\left(\sum_\ell s_\ell^2\right)^2}{\sum_\ell s_\ell^4}=K_xK_yK_t,
\end{equation}
where
\begin{equation}\label{eq:Ka}
K_a=\frac{1+\xi_a^2}{1-\xi_a^2}=\frac12\left(r_a+\frac1{r_a}\right)
\end{equation}
is the Schmidt number in the corresponding dimension, with $a$ taking values $x,y,t$. At low gain, when at most one photon pair is generated per pump pulse, the Schmidt number shows the effective dimensionality of the bipartite entangled state \cite{Gatti12,Horoshko12}. At high gain, when many pairs are generated per pump pulse, another measure, the cooperativity parameter, can be adopted \cite{Christ11}, which is gain-dependent. In our treatment, we will use the Schmidt number $K$ for characterizing the effective number of generated modes, which describes the properties of the squeezing kernel, independent of the pump strength.

Accepting $K$ as a measure of the effective number of modes, we consider only the $K$ modes with the largest singular values, which we call  ``principal modes'', disregarding the rest. We see from Eq.~(\ref{eq:Ks}) that for a Gaussian model under consideration the total Schmidt number is a product of the Schmidt number for spatio-temporal modes, $K_x$, and Schmidt numbers $K_y$ and $K_t$ for the other spatial dimension and the temporal one. For a sufficiently large $K_a$ the corresponding dimension is represented by multiple modes. However, if some of these quantities lies between 1 and 2, the corresponding dimension can be represented by just one mode in the set of principal modes. Let us find the exact ``rounding rule'' for the Schmidt numbers.

We start with just one dimension, putting $K_x=K_y=1$. In this case the ``important'' singular values have the form $\mathcal{N}\xi_x^0\xi_y^0\xi_t^k$ with $k$ running from 0 to $K-1$, the first disregarded singular value being
\begin{equation}\label{sdis}
s_\mathrm{dis}^{1D}=\mathcal{N}\xi_t^K
= \mathcal{N} \left(\frac{K-1}{K+1}\right)^{K/2}
\xrightarrow[K\to\infty]{} \mathcal{N}e^{-1}.
\end{equation}
The convergence in Eq.~(\ref{sdis}) is very fast, and practically takes place at $K>4$. It means that, for a sufficiently high $K$ the first rejected singular value is independent of $K$ and is equal to $\mathcal{N}e^{-1}$. We see here a direct analogy with the exponential decay of some physical quantity by the law $e^{-t/t_c}$, where $t_c$, the time at which the quantity drops to $e^{-1}$ of its initial value, is widely accepted as the characteristic time of the decay process.

Now we analyze two dimensions, putting $K_y=1$ and suggesting that $K_x\in[1,2)$, while $K_t\gg1$. In this case the largest singular values have the form $\mathcal{N}\xi_x^0\xi_y^0\xi_t^k$ or $\mathcal{N}\xi_x^1\xi_y^0\xi_t^k$. The singular values of the first type exhaust all $K$ largest numbers, if the condition $\xi_t^{K-1}>\xi_x$ is satisfied. Writing $K= K_xK_t$ and using the limiting value of Eq.~(\ref{sdis}), we obtain that the last inequality is satisfied for $K_x$ below the critical value $K_0$, being the solution of the equation
\begin{equation}\label{Kcrit}
e^{-K_0} = \sqrt{\frac{K_0-1}{K_0+1}}.
\end{equation}
We find numerically $K_0\approx1.200$. For the values $K_x<K_0$, which corresponds to $\xi_x<0.302$, all principal modes have only zeroth-order mode in the dimension $x$, i.e. the index $i$ in Eq.~(\ref{eq:cl2}) is always zero. For $K_x\ge K_0$ at least one mode with $i=1$ belongs to the set of principal modes. Thus, we have established a rounding rule for the 2D case: if one dimension has a high Schmidt number, while for the second dimension it is less than $K_0$, then this second dimension is single-mode.

\begin{figure}[ht!] 
\includegraphics[width=\linewidth]{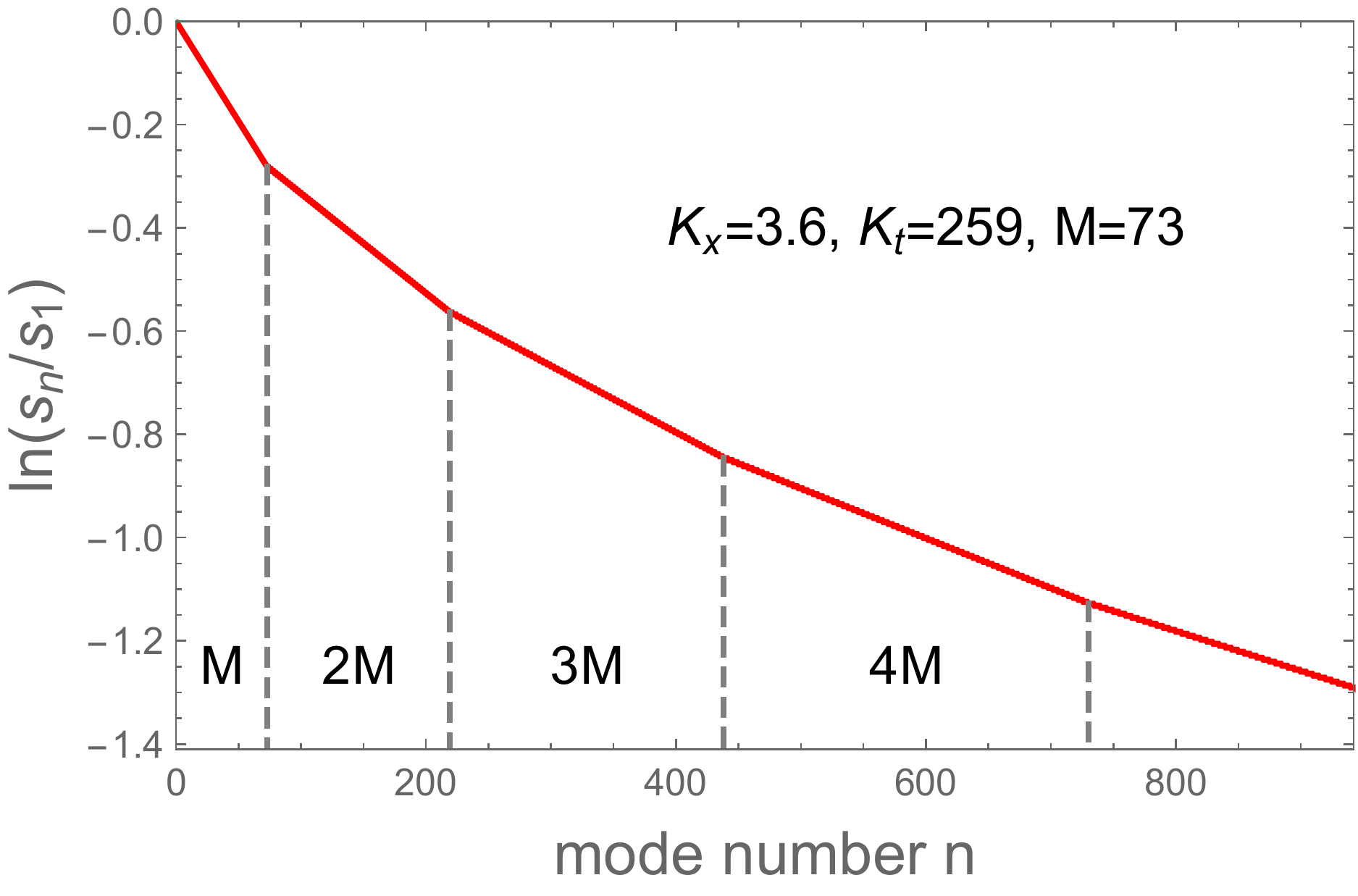}
\caption{Singular values of a Gaussian kernel with $K_y\approx1$, given by Eq.~(\ref{eq:sl}) with $j=0$, as a function of the linear mode number $n$. The dashed vertical lines mark the borders of series, determined by the number of participating spatial modes. The total number of modes in each series is indicated.}
\label{fig:SingVals2D}
\end{figure}

Before treating the 3D case, let us find the first disregarded singular value in the 2D case with $K_t\gg K_x\gg1$. For this purpose we split all principal singular values into series, as shown in Fig.~\ref{fig:SingVals2D}.

The first series contains the values of the form  $\mathcal{N}\xi_x^0\xi_y^0\xi_t^k$, where $k$ runs from 0 to $M-1$ with $M$ being the whole part of $\ln\xi_x/\ln\xi_t$. The second series contains the values of the form  $\mathcal{N}\xi_x^0\xi_y^0\xi_t^k$ with $k$ running from $M$ to $2M-1$, which correspond to the zeroth-order spatial mode, and of the form $\mathcal{N}\xi_x^1\xi_y^0\xi_t^k$ with $k$ running from $0$ to $M-1$, which correspond to the first-order spatial mode. If $\ln\xi_x/\ln\xi_t$ is integer, each value of the first type coincides with some value of the second type, since $\xi_t^M=\xi_x$. We will accept that it is the case for simplicity, though in any practical situation this degeneracy is lifted. Thus, the second series contains $2M$ values corresponding to the spatial modes of orders 0 and 1. In a similar way we find that the third series contains $3M$ values corresponding to the spatial modes of orders 0, 1 and 2, and so on. The total number of values in $J$ series is $MJ(J-1)/2$ and they exhaust all $K$ principal values if approximately $J=\sqrt{2K/M}$. For large $K_a$ the value of $\xi_a$ is well approximated by $\xi_a=e^{-1/K_a}$, as follows from Eq.~(\ref{sdis}). Thus, $M=K_t/K_x$ and $J=\sqrt{2}K_x$. The first disregarded singular value is the first value of the $(J+1)$th series, which is
\begin{equation}\label{sdis2}
s_\mathrm{dis}^{2D}=\mathcal{N}\xi_x^{J}
= \mathcal{N}\xi_x^{\sqrt{2}K_x}
\xrightarrow[K_x\to\infty]{} \mathcal{N}e^{-\sqrt{2}}.
\end{equation}
This result means that in the 2D case the minimal value of $\ln(s_n/s_1)$ for principal modes is $-\sqrt{2}$. In Fig.~\ref{fig:SingVals2D} this value is not reached because $K_x$ is not big enough to attain the limit in Eq.~(\ref{sdis2}).

Now we consider the full 3D case with $K_t\gg K_x\gg1$ and $K_y\in[1,2)$. By the argument of the previous paragraph, $J$ series exhaust $K_xK_yK_t$ values if $J=K_x\sqrt{2K_y}$. Thus, the mode of order 1 in the $y$ dimension does not enter the set of principal modes if the condition $\xi_x^J>\xi_y$ is satisfied, which requires $K_y$ below the value $K_0'$, being the solution of the equation
\begin{equation}\label{KcritY2}
e^{-\sqrt{2K_0'}} = \sqrt{\frac{K_0'-1}{K_0'+1}},
\end{equation}
which is approximately $K_0'\approx1.107$. This corresponds to $\xi_y<0.226$. Thus, we have established a rounding rule for the 3D case: if two dimensions have high Schmidt numbers, while the third dimension has that below $K_0'$, then this third dimension is single-mode.

In the next section we will assume that the condition $K_y<K_0'$ is satisfied, so that the index $j$ in Eq.~(\ref{eq:cl2}) is always equal to zero for the principal modes, which means that the $y$ dimension is single-mode. This assumption is practically reasonable, as explained below, and makes possible a numerical calculation and visualisation of the modal functions.

\section{Simulation of squeezing eigenmodes for BBO \label{sec:sim}}

In this section we apply the developed theory to an example, corresponding to realistic experimental conditions. Two different regimes for the pump are considered  below: one with a rather long and broad pulse, satisfying the requirements for the NPMPA, and a shorter and more focused pump pulse, corresponding to the experimental studies of Ref.~\cite{LaVolpe2020}, which is slightly outside of the area of validity of these requirements.

\subsection{General configuration}
We consider a BBO crystal of length $L=$ \SI{2}{\milli\metre}, illuminated by a vertically polarized pump wave, propagating along the $z$ direction, as described in Sec.~\ref{sec:PDC}. The angle between the optical axis of the crystal and the direction of the pump is $\theta_0=\ang{29.62}$. The central wavelength of the pump is $\lambda_p=\SI{397.5}{\nano\metre}$, which corresponds to the second harmonic of a titanium-sapphire laser. This configuration corresponds to a perfect type-I phase matching for a horizontally polarized ordinary wave at $\lambda_0= \SI{795}{\nano\metre}$ propagating at the angle $\theta_s=\ang{1.8}$ to the $z$ axis, such that $\cos\theta_s=k_p/k_0$. At higher angles perfect phase-matching is attained at higher and lower frequencies, in correspondence with the hyperbolic structure of the phase-matched region, described by Eq.~(\ref{eq:Opm}) (cf. also Fig.~\ref{fig:PDC}).

In this configuration the characteristic spectral and spatial widths of the phase-matching function, defined in Sec.~\ref{subsec:approximations}, are $Q_0=\SI{0.08103}{\per\micro\metre}$ and $\Omega_0=\SI{0.08298}{rad\per\femto\second}$, while the phase-matched transverse wave-vector is $q_d =k_0\sin{\theta_s}= \SI{0.4125}{\per\micro\metre}$. Two mirrors are placed symmetrically in the horizontal plane and select two conjugated segments of the phase-matched cone so that the region $R_s$ is characterized by  $q_{x,\mathrm{min}}=q_d-2\eta_s=\SI{0.3400}{\per\micro\metre}$ and $q_{x,\mathrm{max}}=q_d+2Q_0=\SI{0.5745}{\per\micro\metre}$, which gives the maximal frequency $\Omega_{\mathrm{max}}=\SI{0.4095}{rad\per\femto\second}.$ We assume that the size of the mirrors in the $y$ direction corresponds to the angular size of the pump, so that just one, zeroth-order mode can be considered for this dimension, as discussed in Sec.~\ref{subsec:Schmidt}. This simplifies the analysis and is also reasonable for the chosen method of measurement, discussed in the Introduction (see Fig.~\ref{fig:PDC}). Since we choose two mirrors in the horizontal direction, we can only observe the correlations in the horizontal plain. Two conjugated waves, reflected by the mirrors, arrive at the same point on the beam-splitter only if they lie in the horizontal plane.

We also adopt the general approach of Sec.~\ref{subsec:Schmidt} and limit our consideration to the $K$ principal modes with highest singular values disregarding the rest. Thus, we approximate the JSA, given by Eq.~(\ref{eq:SVD}) by
\begin{equation}\label{eq:SVDapprox}
\mathcal{J}(\vec{q},\Omega,\vec{q}\,{}',\Omega') = h_0(vq_y)h_0(vq_y')
\mathcal{J}_{xt}(q_x,\Omega,q_x',\Omega'),
\end{equation}
where
\begin{equation}\label{eq:Jxt}
\mathcal{J}_{xt}(q_x,\Omega,q_x',\Omega') = \sum_\ell s_\ell C_\ell(q_x,\Omega)D_\ell(q_x',\Omega'),
\end{equation}
is a two-dimensional kernel with the corresponding singular value decomposition into an orthonormal set of functions. It is the kernel $\mathcal{J}_{xt}(q_x,\Omega,q_x',\Omega')$ which will be decomposed numerically below. We see easily from Eq.~(\ref{eq:SVDapprox}) that this kernel can be obtained by putting $q_y=q_y'=0$ in $\mathcal{J}(\vec{q},\Omega,\vec{q}\,{}',\Omega')$. The $K$ left singular functions of $\mathcal{J}(\vec{q},\Omega,\vec{q}\,{}',\Omega')$ with the largest singular values are given by $c_\ell(\vec q,\Omega)=C_\ell(q_x,\Omega)h_0(vq_y)\sqrt{v}$, with a similar expression for the right singular functions. In this way, the problem is reduced to two dimensions, as it was done for photon pairs in Ref.~\cite{Horoshko12}. Such an approach allows for an effective numerical treatment of the problem and a simple visualisation of the modal functions, which are real functions of two arguments.

In the numerical treatment of the problem we follow the way presented in Sec.~\ref{sec:JSA}: We first perform a numerical singular value decomposition of the kernel $\mathcal{J}_{xt}(q_x,\Omega,q_x',\Omega') $ and find the singular functions $C_\ell(q_x,\Omega)$ and $D_\ell(q_x,\Omega)$ with the composite index $\ell=(i,k)$. Note that this kernel corresponds to a real symmetric matrix when discretized in relative coordinates for the signal and idler beams with respect to their perfectly phase-matched points, which reflects the fundamental symmetry of type-I phasematching. As a consequence, the singular value decomposition can be reduced to the spectral decomposition, as discussed in Sec.~\ref{sec:reduction}. Then we build the modal functions of the squeezing eigenmodes $f_\ell^\pm(q_x,\Omega)$ from the singular functions as shown in Eqs.~\eqref{eq:f1}. The squeezing eigenvalues are given by the singular values $s_\ell$, each acquiring  a multiplicity of 2.

The analytical treatment of the problem is presented by Eqs. \eqref{eq:sl}, \eqref{eq:cl2} and \eqref{eq:dl2} with $j=q_y=q_y'=0$, and will be compared with the results of the numerical calculation. Note that in the reduced two-dimensional model the normalization factor reads
\begin{equation}\label{eq:N2}
\mathcal{N}=\pi A_0 \sqrt{(1-\xi_x^2)(1-\xi_t^2)}/(u\tau).
\end{equation}

For better understanding of the experimental conditions and limitations, we replace the Fourier domain characteristics used above with their spatio-temporal analogs. Thus, we characterize the pump pulse by its full width at half maximum, $\tau_p=\sqrt{2\ln2}/\Omega_p$ and its waist $w_p=1/q_p$. Further we introduce two pump-independent characteristics of the setup: time $\tau_0=\sqrt{2\ln2}/\Omega_\mathrm{max}=\SI{2.88}{\femto\second}$ and distance $w_0=1/\eta_s=L\sin\theta_s/(\sqrt{2}\sigma_s)=\SI{27.5}{\micro\meter}$, corresponding to the coherence time and the horizontal walk-off distance of the generated beams respectively. In terms of these parameters, the spatial and temporal analytical Schmidt numbers are given by Eq.~(\ref{eq:Ka}) with $r_x=w_p/w_0$ and $r_t=\mu\tau_p/\tau_0$. The spatio-temporal coupling parameter, Eq.~(\ref{eq:g}), reads
\begin{equation}\label{eq:gbis}
g = N_0\frac{k_0''\mu\sqrt{w_p L}}{\pi\tau_p \sqrt{\sin\theta_s}}\Omega_\mathrm{max},
\end{equation}
where the numerical factor is $N_0=\pi\sqrt{\ln2/(2^{3/2}\sigma_s)}\approx1.23$. To obtain a high degree of squeezing in an experiment, one needs a high pump intensity, which can be reached using ultrashort and focused pump pulses. Equation~(\ref{eq:gbis}) shows that for a high spatio-temporal coupling, the pump pulses have to be as short as possible within the limits of used approximations, but not very focused.

The justification of the analytical Gaussian model relies on the applicability of the NPMPA, which requires that the pump parameters surpass the values $\tau_\mathrm{NPMPA}=141$~fs and $w_\mathrm{NPMPA}=21.4$~$\mu$m, as found in Appendix~\ref{appendix:npmpa}.

\subsection{Long and wide pump pulse \label{subsec:long}}

\begin{figure}[ht!] 
\center{\includegraphics[width=1\linewidth]{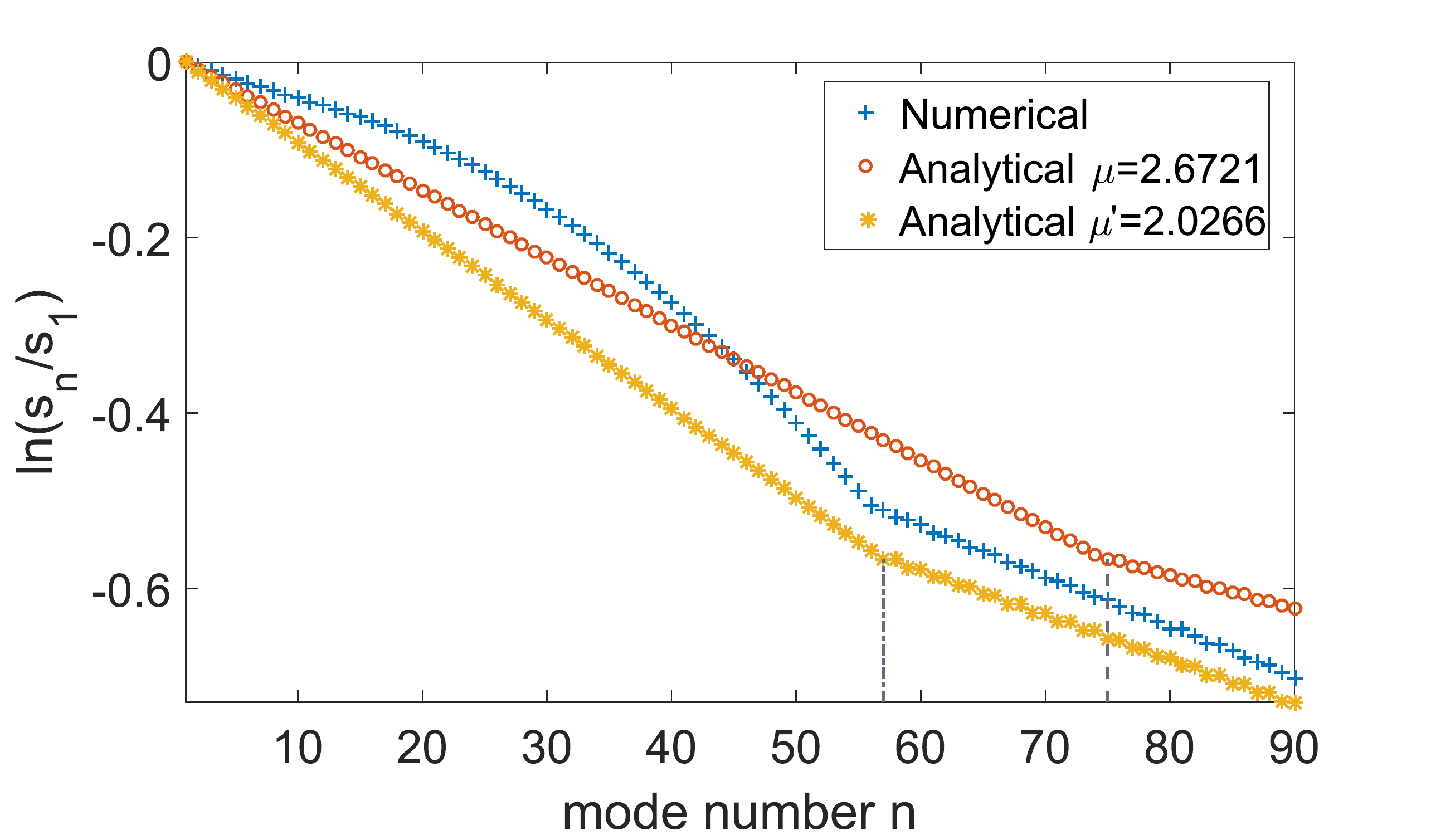}}
\caption{Singular values for the case of  a \SI{280}{\femto\second} duration and \SI{100}{\micro\metre} waist pump. The vertical dashed lines separate series of values corresponding to a given number of spatial modes, as discussed in Sec.~\ref{subsec:Schmidt}. The yellow stars represent the analytical solution where the fitting parameter is chosen to adjust the analytical bending point $M$ to the numerical one.}.
\label{fig:EigvalsLonger}
\end{figure}

First we consider a pump pulse, which satisfies the requirements of the NPMPA, having a duration $\tau_p=$ \SI{280}{\femto\second}, and a waist $w_p=$ \SI{100}{\micro\metre}. The fitting parameter $\mu$ enters the definition of $\tau$ and therefore changes the scale of the analytical modal functions in the  $\Omega$ direction. We optimize it in order to have the highest average overlap between the analytical and numerical shapes of the first six modes. For the considered pump size we find $\mu=2.6721$. In Fig.~\ref{fig:EigvalsLonger} we report the numerical and analytical singular values in a logarithmic scale as functions of the linear mode number $n$, which is obtained from the composite mode number $\ell=(i,k)$ sorting the modes in the descending order of their singular values.

We find the analytical values $K_x=1.96$ and $K_t=129$, giving the total number of modes $K=K_xK_t=253$. We see that the number of modes for a long and weakly focused pump may be very high. An analysis of a similar configuration in the low-gain regime with even larger pump size gave the number of modes above $10^5$ \cite{Gatti12}. Only the 90 highest singular values are shown in Fig.~\ref{fig:EigvalsLonger}. Since $\xi_t=0.9923$ and $\xi_x=0.5688$, we can observe the first change of slope in the analytical curve after the value $M=\ln{\xi_x}/\ln{\xi_t}=73.05$, as explained in section \ref{subsec:Schmidt}. The numerical singular values do not follow the piecewise-linear shape of the analytical Gaussian model, but have a similar change of slope after $n=57$. Changing the fitting parameter to $\mu'=2.0266$, we could adjust the bending point of the analytical singular values to that of the numerical ones, as shown by the yellow stars in Fig.~\ref{fig:EigvalsLonger}. However, the numerical and analytical modal functions would in this case have a lower overlap. Thus, we keep in the following the original fitting parameter, optimized for the modal functions.

\begin{figure*}[ht] 
\center{\includegraphics[width=\textwidth]{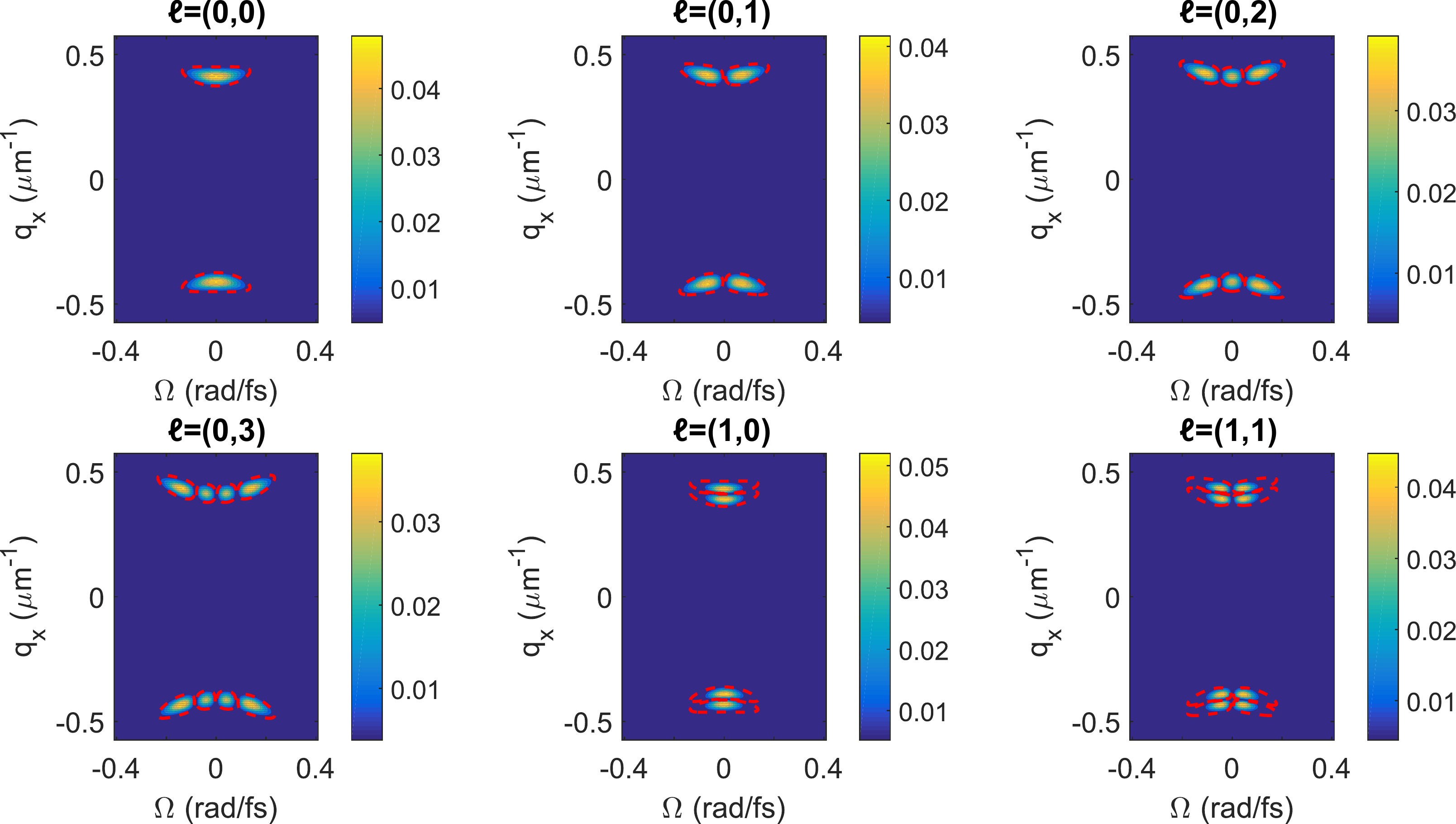}}
\caption{Spectral modal functions of the squeezing eigenmodes for the case of a \SI{280}{\femto\second} duration and \SI{100}{\micro\metre} waist pump. The modulus of the numerical modal function is shown with the color map while the 2$\sigma$ area of the analytical solution is marked by the red dashed line. The mode with $\ell=(1,0)$ has a linear number $n=57$ and corresponds to the ``bending point'' of Fig.~\ref{fig:EigvalsLonger}. Before this point only the zero-order spatial modes exist. After this point the first-order spatial modes have singular values similar to the zero-order spatial modes. The overlap between the analytical solution and the numerical simulation is 1.00 for modes n=1 to n=4, for n=57 it is 0.95 and for n=58 it is 0.87.}
\label{fig:spatiotemp_modes_1}
\end{figure*}

The spectral modal functions $f_{\ell}^+(q_x,\Omega)$ of the squeezing eigenmodes are shown in Fig.~\ref{fig:spatiotemp_modes_1}. The functions $f_{\ell}^-(q_x,\Omega)$ differ from them by the sign of the idler part.

The modal function of a squeezing eigenmode is a combination of two singular functions for the signal ($q_x>0$) and the idler ($q_x<0$) beams. We see that the first four shown singular functions indeed represent the same spatio-temporal functions $h_0\left(uq_x \mp uq_d \mp g\tau^2\Omega^2\right)$, given by Eqs.~\eqref{eq:cl2} and \eqref{eq:dl2}, which are modulated in the $\Omega$ dimension by functions $h_k(\tau\Omega)$. At higher orders, $\ell=(1,0)$ and $(1,1)$, we see the function $h_1\left(uq_x \mp uq_d \mp g\tau^2\Omega^2\right)$ modulated in the same way in the $\Omega$ dimension. As can be seen, there is almost a perfect agreement between the analytical solution and the numerical simulation, though the overlap reduces for higher order modes.

We see a clear spatio-temporal structure in all the modes and the effect is even more clear for higher order modes. Coupling of the spatial and temporal degrees of freedom is represented by the ``curved'' shape of the modal functions, which cannot be represented as a product of a function of $q_x$ and a function of $\Omega$. The Gaussian modeling allows us to introduce a measure of this coupling. Let us analyze the zeroth-order spatial mode with the signal modal function
\begin{equation}\label{eq:C00}
C_{0k}(q_x,\Omega)=h_0\left(uq_x - uq_d - g\tau^2\Omega^2\right)h_k(\tau\Omega)\sqrt{u\tau}.
\end{equation}
The degree of spatio-temporal coupling can be characterized by the ratio of its vertical deflection from $q_d$ at the edges, let us denote it by $\Delta q$, to its half-width in the center ($\Omega=0$), which can be estimated by two standard deviations, $\delta q=2/u$. The edge of the horizontal half-width can be found from the following consideration. All $k$ zeros of the Hermite-Gauss function $h_k(x)$ lie between $-\sqrt{4k+3}$ and  $\sqrt{4k+3}$ \cite{AbramowitzStegun}. Thus, the edge frequency of $h_k(\tau\Omega)$ can be estimated as $\Omega=\sqrt{4k+3}/\tau$. At this frequency the modal function is maximal at $q_x=q_d+(4k+3)g/u$, i.e. the vertical deflection is $\Delta q=(4k+3)g/u$. Thus, the coupling is determined by the value $\Delta q/\delta q=(2k+\frac32)g$. In the considered case $g\approx0.28$, which corresponds to a well-seen coupling at $k=3$, where $\Delta q/\delta q\approx2$. The coupling is more and more pronounced with the growth of the temporal mode number $k$, because the width of the function $h_k(\tau\Omega)$ grows.

The spatio-temporal modal functions can be obtained by a double Fourier transform of the spectral functions, defined by Eq.~(\ref{eq:f1}):
\begin{equation}
    F_\ell^\pm(x,t)=\frac1{2\pi}\int f_\ell^\pm(q_x,\Omega)e^{i(q_xx-\Omega t)}
    dq_xd\Omega.
\end{equation}
An analytic expression can be obtained for the zeroth-order temporal modes:
\begin{equation}
F_{i0}^+(x,t)= \frac{\sqrt{2}h_i(x/u)}{\sqrt{\pi^{1/2}u\tau}} \mathrm{Re}\left(\frac{e^{iq_dx-\frac12(t/\tau)^2/(1-2igx/u)}}{\sqrt{1-2igx/u}}\right)
\end{equation}
and a similar expression for $F_{i0}^-(x,t)$ with the real part replaced by the imaginary one. Note that these functions are normalized to unity in the limit $q_du\gg1$.

\begin{figure}[ht!] 
\includegraphics[width=0.8\linewidth]{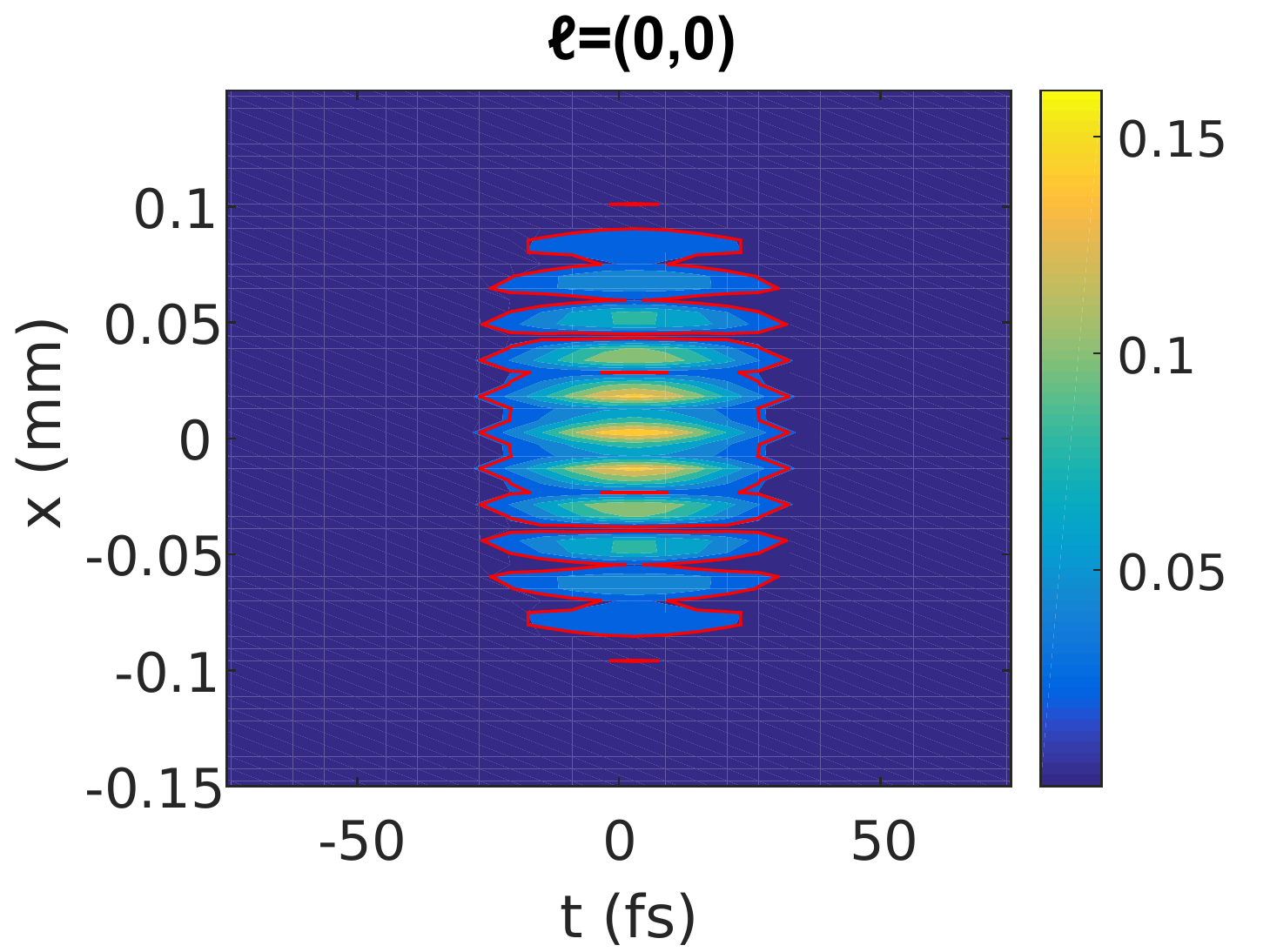}
\includegraphics[width=0.8\linewidth]{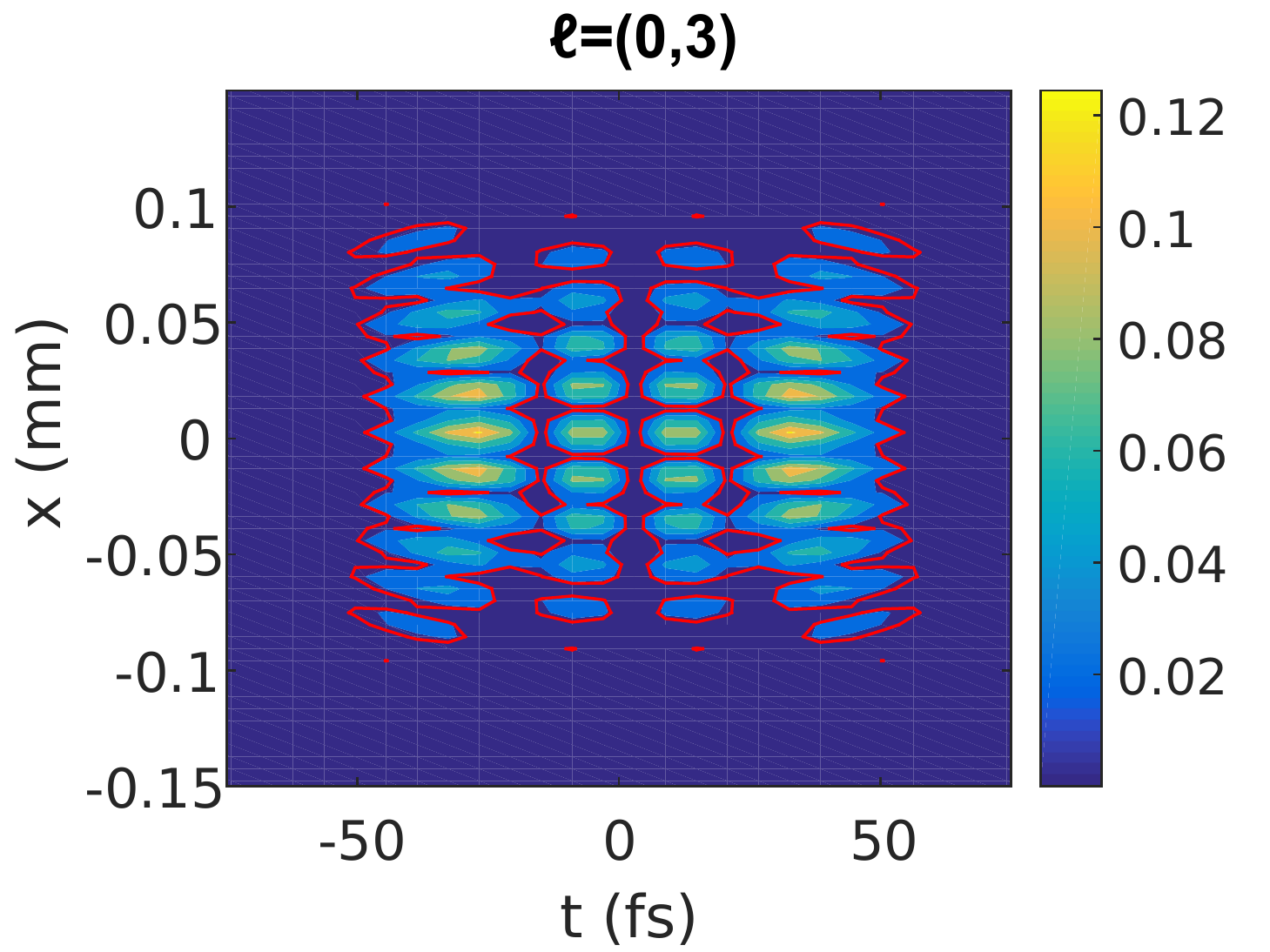}
\caption{Spatio-temporal modal functions, calculated as numerical Fourier transforms of the spectral modal functions for the case of a \SI{280}{\femto\second} duration and \SI{100}{\micro\metre} waist pump. The modulus of the simulated modal function is shown by the color map. The contour line (solid red line) is the 2$\sigma$ area of the analytical solution. The non-factoring spatio-temporal structure is well visible for the mode $\ell=(0,3)$. }
\label{fig:SwitchingModesLonger}
\end{figure}

The spatio-temporal modal functions are shown in Fig.~\ref{fig:SwitchingModesLonger} as numerical Fourier transforms of the corresponding spectral modes shown in Fig.~\ref{fig:spatiotemp_modes_1}. We recall that these modes correspond to the field formed at the output face of the nonlinear crystal, linearly backprojected onto the crystal center. These distributions can be transmitted to the camera by a proper imaging system. The non-factoring structure of the modes is a direct consequence of the spatio-temporal coupling determined by the parameter $g$.

\subsection{Short and focused pump pulse}

As the second example we consider a shorter pump pulse with the FWHM duration $\tau_p=$ \SI{128}{\femto\second} focused at the center of the crystal to the waist $w_p=$ \SI{49}{\micro\metre}, which corresponds to a recent experiment reported in Ref.~\cite{LaVolpe2020}. The fitting parameter chosen to maximize the overlap of the first six modal functions is $\mu=0.92421$.

\begin{figure}[ht!] 
\center{\includegraphics[width=1\linewidth]{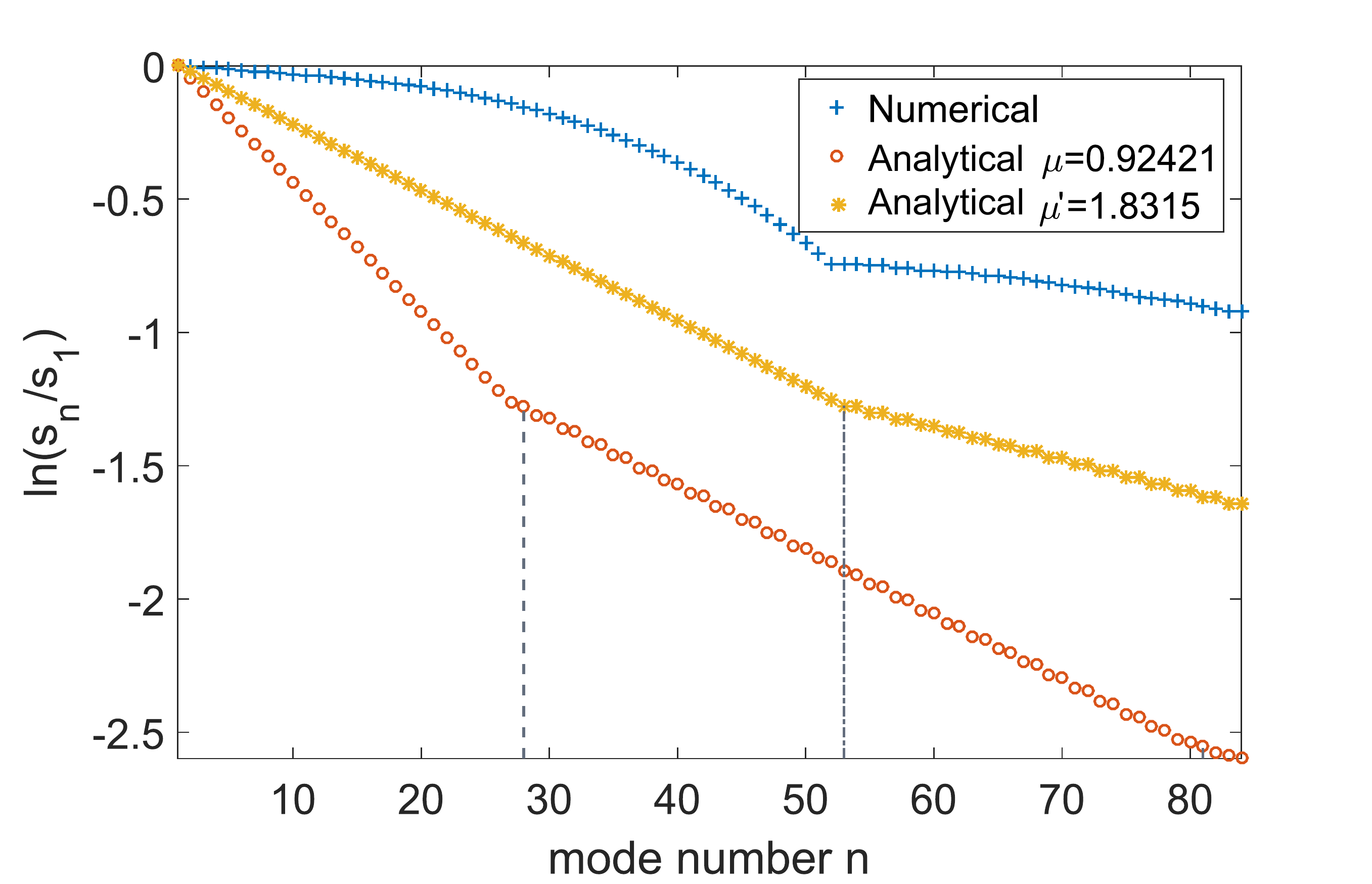}}
\caption{Singular values for the case of shorter and more focused pump, corresponding to the experimental parameters of Ref.~\cite{LaVolpe2020}. The vertical dashed lines separate series of values corresponding to a given number of spatial modes, as discussed in Sec.~\ref{subsec:Schmidt}. The yellow stars represent the analytical solution where the fitting parameter is chosen in order to adjust the analytical bending point $M$ to the numerical one.}
\label{fig:EigvalsShorter}
\end{figure}

For these pump parameters, the analytical Gaussian modeling gives the Schmidt number in the $x$ dimension $K_x=1.17$. This value is lower than the critical value $K_0=1.200$,  found in Sec.~\ref{subsec:Schmidt}. It means that no modes with $i=1$ appear in the set of principal modes. The Schmidt number in the $t$ dimension is $K_t=20.5$, which gives the total number of modes $K=K_xK_t=24$. The numerically simulated singular values are shown in Fig.~\ref{fig:EigvalsShorter}. In this case $\xi_t=0.9524$ and $\xi_x=0.2813$, and we can observe the first change of slope in the analytical curve after the value $M=\ln{\xi_x}/\ln{\xi_t}=26$. As in the previous section, we see that a different value of $\mu'=1.8315$ would allow us to adjust the structure of the analytic singular values to that of the numerical ones. However, it would lead again to a poorer overlap of the modal functions.

\begin{figure*}[ht!] 
\includegraphics[width=.24\linewidth]{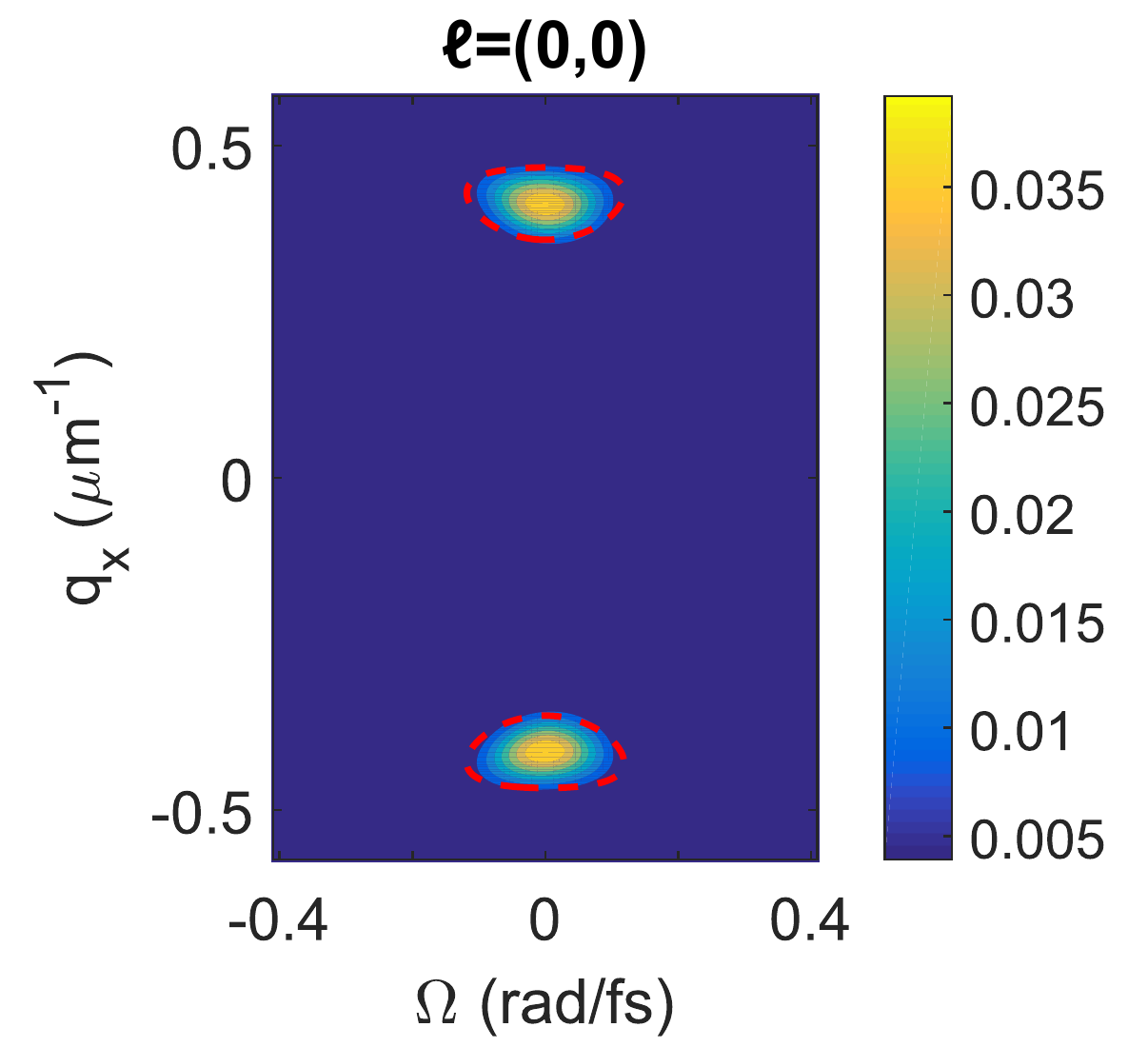}
\includegraphics[width=.24\linewidth]{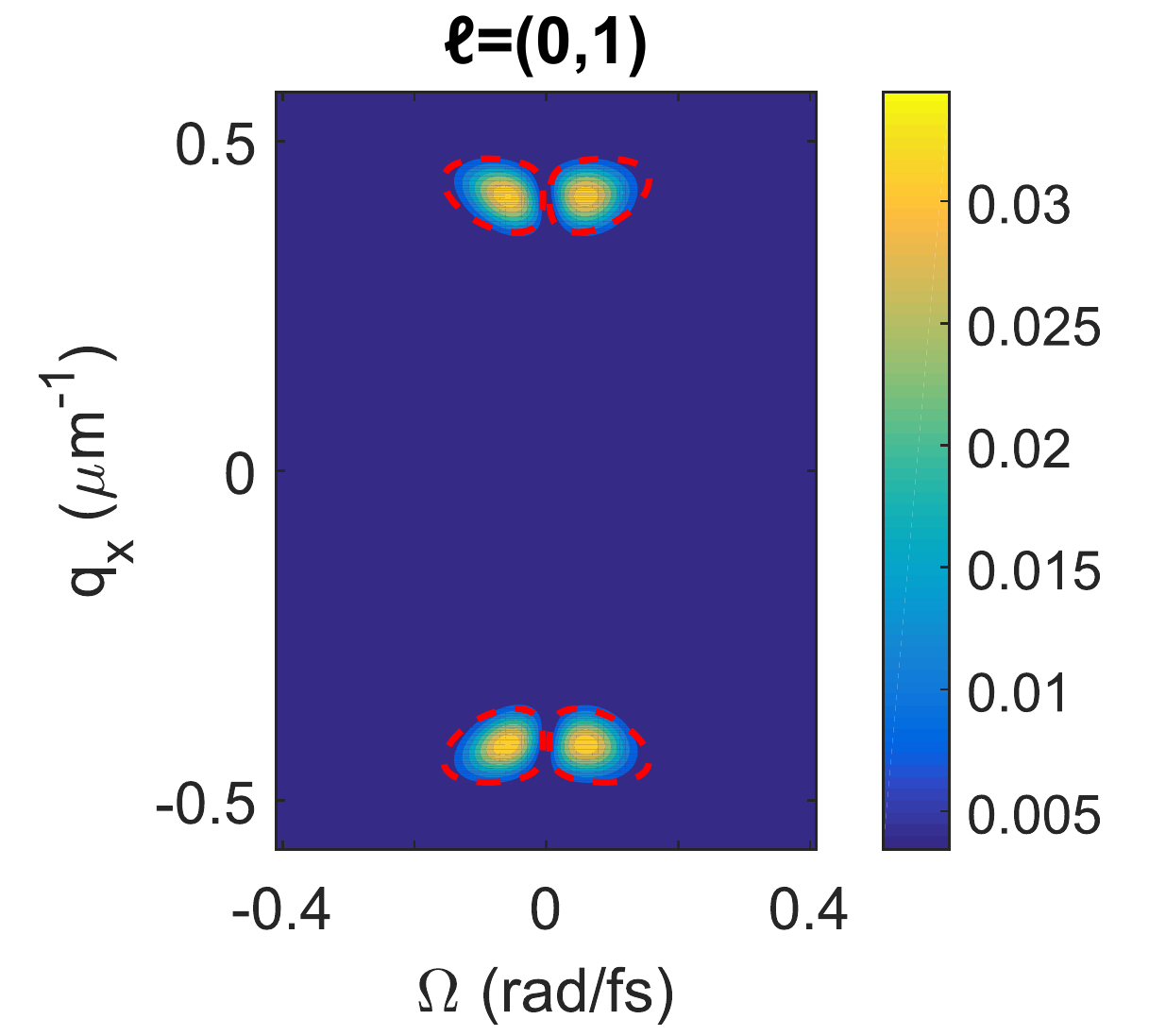}
\includegraphics[width=.24\linewidth]{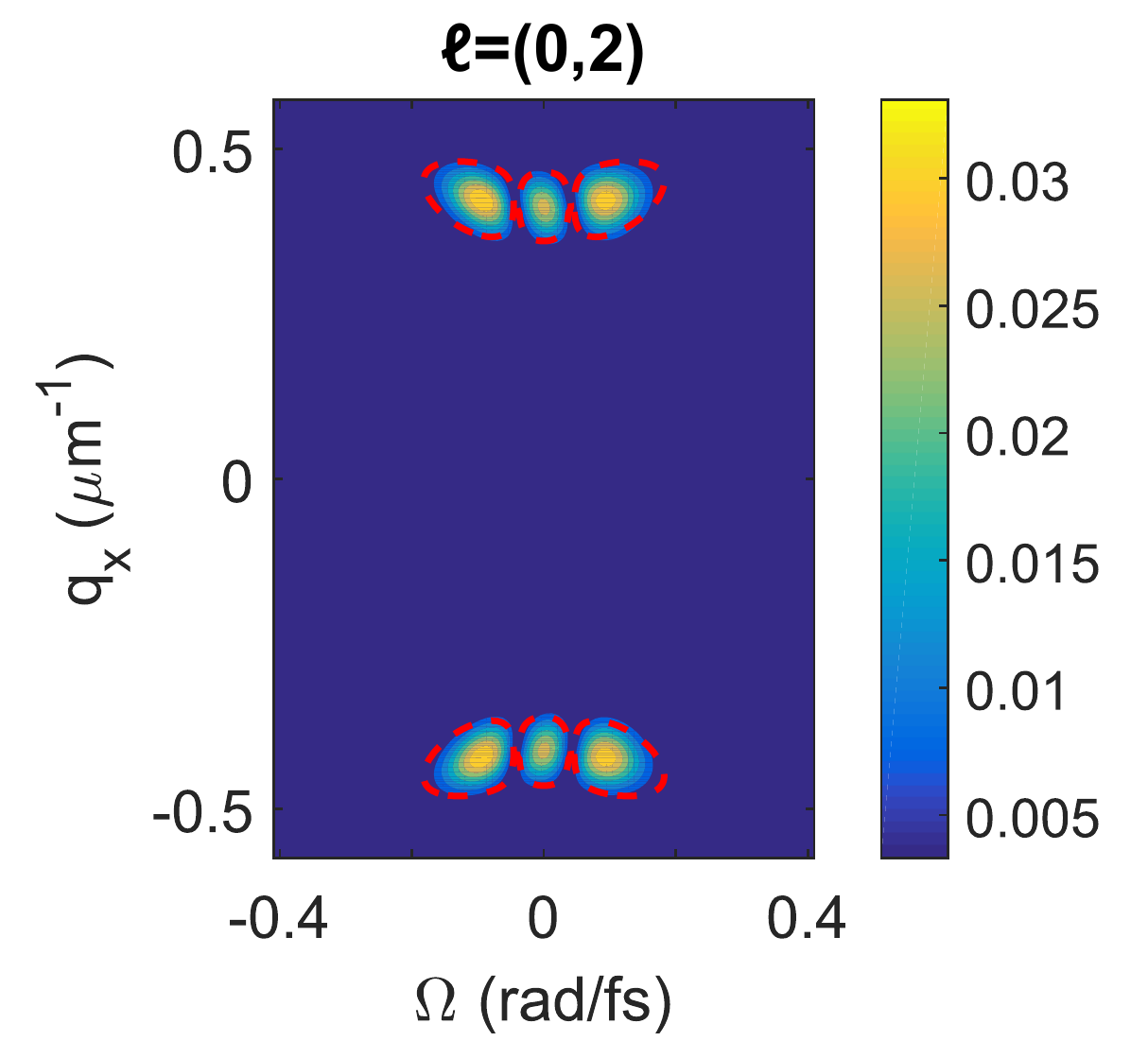}
\includegraphics[width=.24\linewidth]{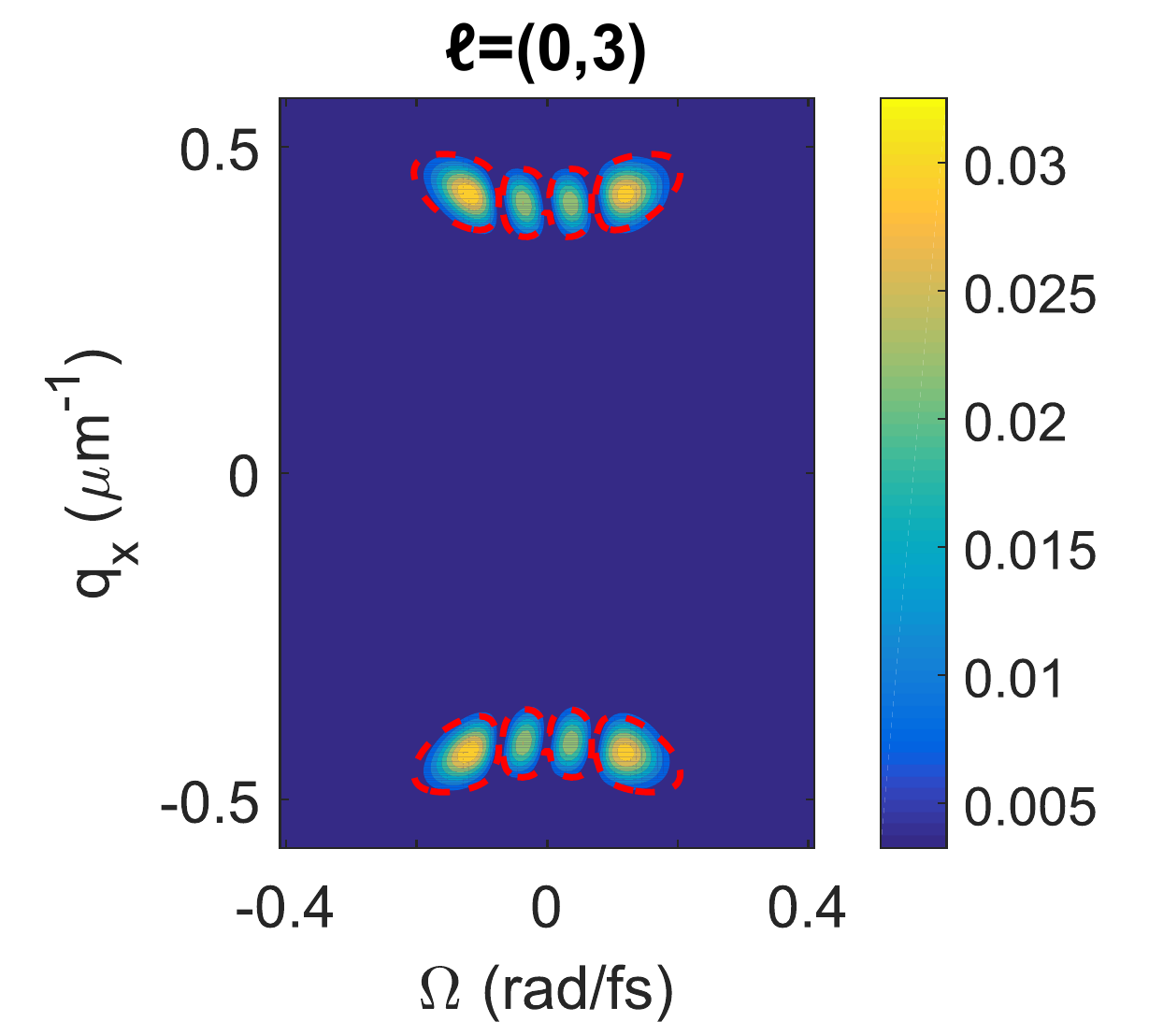}
\caption{Spectral modal functions of the squeezing eigenmodes for the case of a \SI{128}{\femto\second} duration and \SI{49}{\micro\metre} waist pump. The modulus of the numerically simulated modal function is shown with the color map while the 2$\sigma$ area of the analytical solution is marked by the red dashed line. The overlaps between the analytical solution and the corresponding simulation are $(0.988, 0.986, 0.984, 0.982)$ from left to right.}
\label{fig:ModesShorter}
\end{figure*}

The spectral modal functions for the first four modes are shown in Fig.~\ref{fig:ModesShorter}, where we again see a very good correspondence between the numerical and the analytical solutions. This is remarkable, because the size of the pump pulse in the considered case does not satisfy the conditions of applicability of NPMPA, formulated in Appendix~\ref{appendix:npmpa}. It means that application of this approximation may be useful even at a shorter and more focused pump. On the other hand, comparing Fig.~\ref{fig:EigvalsShorter} to Fig.~\ref{fig:EigvalsLonger}, we conclude that the correspondence between the analytical and the numerical singular values becomes worse outside the conditions of NPMPA.

The coupling between $q_x$ and $\Omega$ is less pronounced in the considered case, which is explained by a rather low value of $g=0.16$. The first-order spatial mode with $\ell=(1,0)$ appears at $n=53$, as is clear from the bending point in Fig.~\ref{fig:EigvalsShorter}. This mode does not belong to the set of principal modes, and we can say that the considered case is spatially single-mode.

\begin{figure}[ht!] 
\includegraphics[width=.8\linewidth]{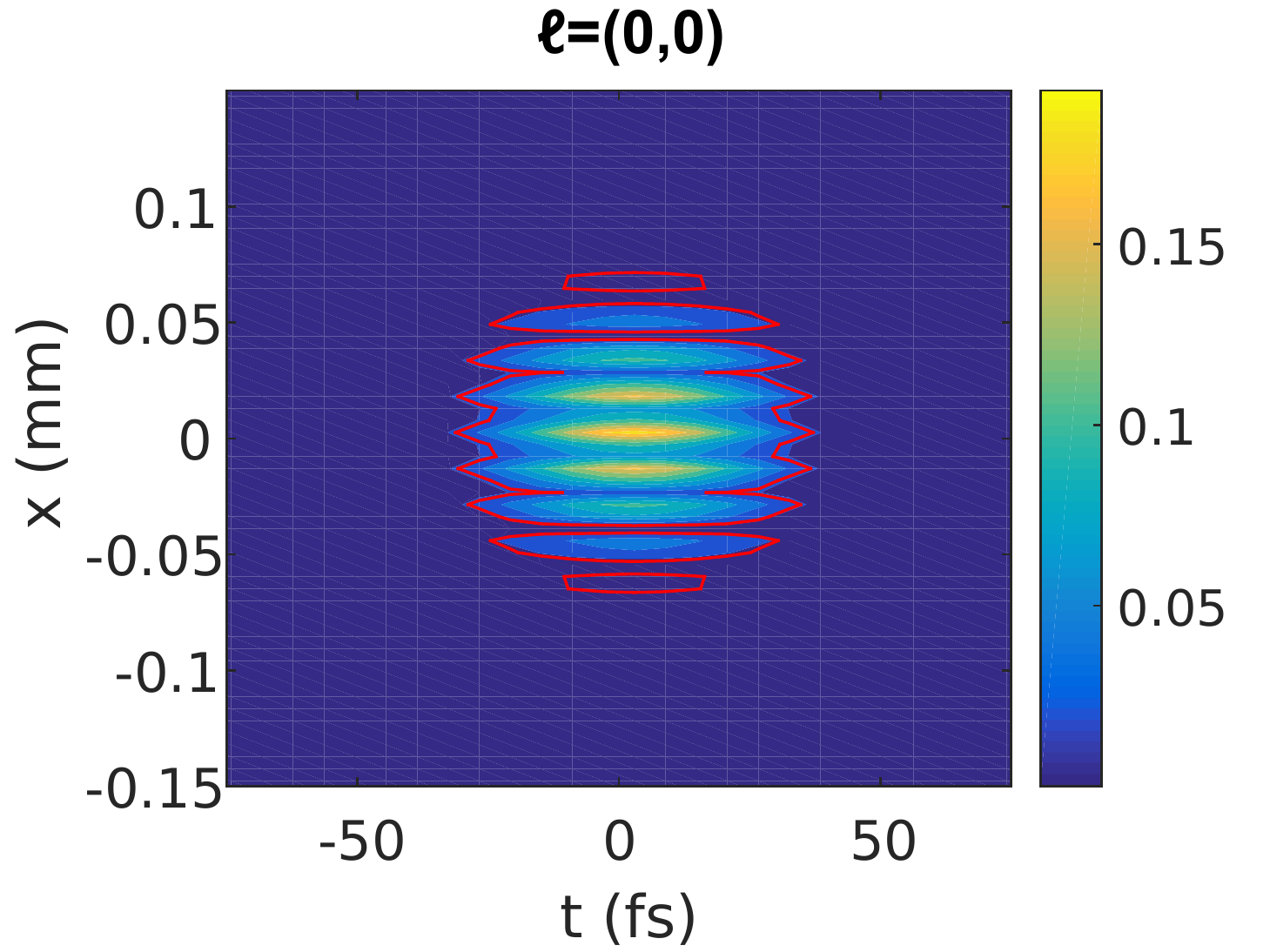}
\includegraphics[width=.8\linewidth]{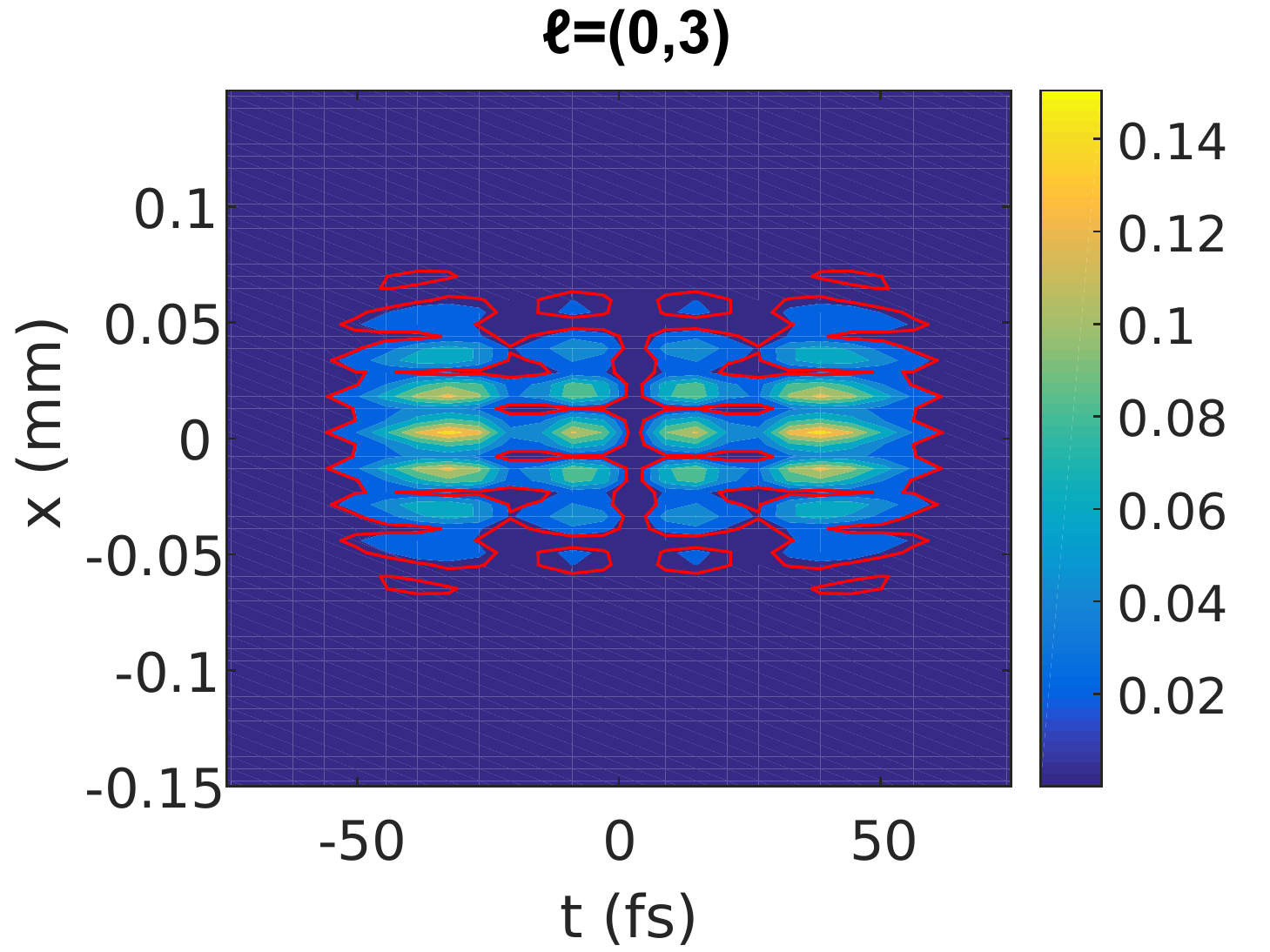}
\caption{Spatio-temporal modal functions, calculated as numerical Fourier transforms of the spectral modal functions for the case of a \SI{128}{\femto\second} duration and \SI{49}{\micro\metre} waist pump. The modulus of the numerically simulated modal function is shown by the color map.  The contour line (solid red line) is the 2$\sigma$ area of the analytical solution. }
\label{fig:spatiotemp modes}
\end{figure}

The spatio-temporal modal functions are presented in Fig.~\ref{fig:spatiotemp modes}. In the considered regime the non-factoring structure of these functions is less pronounced. However, some signature of the coupling between the spatial and temporal degrees of freedom was observed in this regime in the experiment \cite{LaVolpe2020}. As shown in Eq.~\eqref{eq:gbis} the degree of spatio-temporal coupling of the modes can be enhanced or reduced by carefully optimizing the ratio between the pump waist and the pump pulse duration.

Observation of squeezing requires a recombination of the signal and idler beams on a beamsplitter with a subsequent homodyne detection of one of its outputs, as discussed in Introduction. For a mode-selective measurement, the local oscillator should be shaped both in space and time to match the chosen mode. We see that in realistic cases studied in this and the previous sections the temporal mode number can be of the order of 100, while the spatial mode number is not higher than 1. The resolution of state-of-the-art pulse shapers allows one to work with such a high number of modes \cite{Brecht15}.

\section{Conclusions \label{sec:Conclusions}}

We have studied the spatio-temporal structure of entangled beams of light generated in a single-pass noncollinear OPA. The main result of our treatment is the analytical Gaussian model of the squeezing kernel in curvilinear coordinates, which allows one to obtain analytical expressions for the modal functions of the squeezing eigenmodes and the corresponding singular values. The modal functions obtained in this way are non-factoring in space and time and are in an excellent correspondence with the numerically found eigenfunctions of the squeezing kernel. The structure of the eigenvalues in the general 3D case is analyzed in detail and the conditions for essentially 1D and 2D cases are found. Also, the analytical model gives an expression for the degree of spatio-temporal coupling, and, in particular, predicts, that this degree is growing with the pump waist and decreasing with the pump duration. This model gives a simple rule for estimating the strength of spatio-temporal coupling.


Applications of the developed theory go far beyond the cluster state quantum computation mentioned in the Introduction. Recombination of frequency-degenerate entangled EPR beams on a beamsplitter, shown in Fig.~\ref{fig:PDC}, creates an SU(1,1) nonlinear interferometer, which is highly prospective for phase estimation at the Heisenberg limit \cite{Ou92apb,Ou20} and for optical coherence tomography \cite{Machado20}. In classical optics, the noncollinear geometry is widely used in OPAs for increasing the amplification bandwidth \cite{Dubietis06}, and studying the modal structure of the signal and idler beams is important for proper mode-matching. Besides, in the low-gain regime, the quantum state of the generated photon pairs is determined by the modal structure of the PDC radiation \cite{Fabre20}, which is the same as in the first Magnus approximation of the high-gain regime, analyzed here. In this context, our study of the spatiotemporal coupling can be helpful for avoiding such a coupling where it is undesirable. A precise definition of the temporal modes is crucial for applications in quantum information science \cite{Brecht15} and photonic quantum sensing \cite{Pirandola18}.

\begin{acknowledgments}
This work was supported by the network QuantERA of the European Union's Horizon 2020 research and innovation programme under project ``Quantum information and communication with high-dimensional encoding'' (QuICHE) and also by H2020 Future and Emerging Technologies (665148), Agence Nationale de la Recherche (ANR-14-CE32-0019), European Research Council (820079).
\end{acknowledgments}

\appendix

\section{Wave-vector of the pump}\label{appendix:nex}
Here we show how the longitudinal component of the pump wave-vector $k_{pz}(\vec q,\Omega)$, defined by Eq.~(\ref{eq:kpz}) can be found from the ordinary and extraordinary refractive indices of a uniaxial crystal, denoted $n_o(\omega)$ and $n_e(\omega)$ respectively. Let us denote the unit vectors along the directions $x,y,z$ as $\vec n_x,\vec n_y,\vec n_z$ respectively. The pump is composed of many monochromatic plane waves, each having a wave-vector $\vec k_{p}(\vec q,\Omega)=k_{pz}(\vec q,\Omega)\vec n_z+\vec q$. The refractive index along the direction of propagation of such a wave is \cite{BoydBook}
\begin{equation}\label{eq:np}
n_{p}(\vec{q},\omega)=\sqrt{\frac{1}{\frac{\sin^2\theta(\vec{q},\omega)}{n_e(\omega)^2} +\frac{\cos^2\theta(\vec{q},\omega)}{n_o(\omega)^2}}},
\end{equation}
where $\theta(\vec{q},\omega)$ is the angle between the vector $\vec k_{p}(\vec q,\Omega)$ and the optical axis of the crystal. The unit vector in the direction of the optical axis is $\vec n_{OA} = \cos{\theta_0}\vec n_z + \sin{\theta_0}\vec n_y$, and the angle of interest is determined by the relation
\begin{equation}\label{eq:theta}
\cos\theta(\vec{q},\omega) = \frac{\vec n_{OA}\vec k_{p}(\vec q,\Omega)}{k_{p}(\vec q,\Omega)}
= \frac{k_{pz}(\vec q,\Omega)\cos{\theta_0}+q_y\sin{\theta_0}}{k_{p}(\vec q,\Omega)},
\end{equation}
where $k_{p}(\vec q,\Omega)=|\vec k_{p}(\vec q,\Omega)|$. The latter can be written as
\begin{equation}\label{eq:kp}
k_{p}(\vec q,\Omega)=n_{p}(\vec{q},\omega_p+\Omega)(\omega_p+\Omega)/c,
\end{equation}
and Eq.~(\ref{eq:kpz}) can be rewritten as
\begin{equation}\label{eq:kp2}
k_{p}^2(\vec q,\Omega)=k_{pz}^2(\vec q,\Omega)+q_x^2+q_y^2.
\end{equation}

Equations (\ref{eq:np}), (\ref{eq:theta}), (\ref{eq:kp}), (\ref{eq:kp2}) represent a system of four equations with four unknown functions, which can be resolved with respect to $k_{pz}(\vec q,\Omega)$. Squaring both sides of Eq.~(\ref{eq:np}) and both sides of Eq.~(\ref{eq:kp}), and excluding $n_{p}^2(\vec{q},\omega)$ from the resulting equations, we obtain
\begin{equation}\label{eq:excl1}
k_{p}^2(\vec q,\Omega)\left(\frac{1-\cos^2\theta(\vec{q},\omega)}{n_e(\omega)^2} +\frac{\cos^2\theta(\vec{q},\omega)}{n_o(\omega)^2}\right) = \frac{\omega^2}{c^2}.
\end{equation}
Squaring both sides of Eq.~(\ref{eq:theta}) and substituting the result into Eq.~(\ref{eq:excl1}), we obtain
\begin{eqnarray}\label{eq:excl2}
\frac{k_{pz}^2(\vec q,\Omega)+q_x^2+q_y^2}{n_e(\omega)^2} &+&\left(\frac1{n_o(\omega)^2}-\frac1{n_e(\omega)^2}\right) \\\nonumber
&\times&\left(k_{pz}(\vec q,\Omega)\cos{\theta_0}+q_y\sin{\theta_0}\right)^2 = \frac{\omega^2}{c^2}.
\end{eqnarray}
where we have also excluded $k_{p}^2(\vec q,\Omega)$ by employing Eq.~(\ref{eq:kp2}).

Equation (\ref{eq:excl2}) is a quadratic equation with respect to $k_{pz}(\vec q,\Omega)$. Its positive root is
\begin{eqnarray}\label{eq:root}
k_{pz}(\vec q,\Omega)&=& \left(\frac{n_z(\omega)^2}{n_o(\omega)^2}-\frac{n_z(\omega)^2}{n_e(\omega)^2}\right)q_y\sin{\theta_0}\cos{\theta_0} \\\nonumber
&+&\sqrt{\frac{n_z(\omega)^2\omega^2}{c^2}-\frac{n_z(\omega)^2q_x^2}{n_e(\omega)^2}-\frac{n_z(\omega)^4q_y^2}{n_o(\omega)^2n_e(\omega)^2}},
\end{eqnarray}
where we have defined the extraordinary refractive index in the $z$ direction
\begin{equation}\label{eq:nz}
n_{z}(\omega)=n_{p}(0,\omega)=\sqrt{\frac{1}{\frac{\sin^2\theta_0}{n_e(\omega)^2} +\frac{\cos^2\theta_0}{n_o(\omega)^2}}}.
\end{equation}

Equation (\ref{eq:root}) is used in this article for numerical simulation of the squeezing kernel with $n_o(\omega)$ and $n_e(\omega)$ given by the Sellmeier equations for BBO crystal \cite{Eimerl87}.

\section{Limits of the NPMPA}\label{appendix:npmpa}
The nearly plane-wave and monochromatic approximation, introduced in Sec.~\ref{subsec:approximations}, is valid for a pump pulse which is long and wide enough. The exact conditions for the pulse being considered long and wide are established in Ref.~\cite{Caspani10}. However, in this reference the conditions are formulated in a form of inequalities which are to be satisfied by all spatio-spectral components of the downconverted light. Here we apply the other basic approximation, the PQDA also introduced in Sec.~\ref{subsec:approximations}, and obtain explicit expressions for the characteristic time and distance, which should be surpassed by the corresponding spatio-temporal dimensions of the pump pulse.

We start with expanding the phase mismatch, defined by Eq.~(\ref{eq:Delta}), in the Taylor series up to the second order in $(\vec{q}_-,\Omega_-)$, as required by PQDA, and to the first order in $(\vec{q}_+,\Omega_+)$, which gives the lowest-order terms disregarded in NPMPA:
\begin{eqnarray}\label{eq:Delta3}
&&\Delta\left(\frac{\vec{q}_+}2+\vec{q}_-, \frac{\Omega_+}2+\Omega_-, \frac{\vec{q}_+}2-\vec{q}_-, \frac{\Omega_+}2-\Omega_-\right) \\\nonumber
&&\approx 2k_0-k_p + k_0''\Omega_-^2 - \frac1{k_0}\left(q_{x-}^2+q_{y-}^2\right)\\\nonumber
&&+ \frac{\tau_\mathrm{wo}\Omega_+}L  -\frac{x_\mathrm{wo}q_{x+}}L -\frac{y_\mathrm{wo}q_{y+}}L -\rho_pq_{y+},
\end{eqnarray}
where we have used Eq.~(\ref{eq:root}) and the notation $\vec{q}_\pm=(q_{x\pm},q_{y\pm})$. We recognize in the first four terms in the right hand side of Eq.~(\ref{eq:Delta3}) the PQDA and NPMPA phase mismatch $\Delta(\vec{q_-},\Omega_-,-\vec{q_-},-\Omega_-)$, introduced by Eq.~(\ref{eq:Delta2}). The other terms represent linear deflections from this form, which are supposed to be small. Here
\begin{equation}
\tau_\mathrm{wo} = \left|k_0'-k_p'+k_0'\frac{q_{x-}^2+q_{y-}^2}{2k_0^2}\right|L = \left|\frac{L}{v\cos\phi}-\frac{L}{v_p}\right|
\end{equation}
is the walk-off time for the pump propagating along the $z$ axis at the group velocity $v_p=1/k_p'$ and the pair of signal-idler components at the degenerate frequency with the transverse wave-vectors $\vec{q}_-$ and $-\vec{q}_-$ respectively, propagating each at the group velocity $v=1/k_0'$ at the angle $\phi=\arccos(k_z(\vec{q}_-,0)/k_0)$ to the $z$ axis.

In a similar way we define the walk-off distance in the $x$ direction between the signal component at $(q_{x-},0,\Omega_-)$ and the $z$ axis
\begin{equation}
d_x^{(s)} = \frac{q_{x-}}{k_0}\left(1-\frac{k_0'\Omega_-}{k_0}\right)L = L\tan\psi,
\end{equation}
where $\psi=\arcsin[q_{x-}/k(\Omega_-)]$ is the angle between the component $(q_{x-},0,\Omega_-)$ and the $z$ axis, which is so small in PQDA that $\sin\psi\approx\tan\psi$. The idler twin of this signal component is deflected in the opposite direction by
\begin{equation}
d_x^{(i)} = -\frac{q_{x-}}{k_0}\left(1+\frac{k_0'\Omega_-}{k_0}\right)L,
\end{equation}
The average deflection of the signal-idler pair is
\begin{equation}\label{xwo}
x_\mathrm{wo}=\frac{d_x^{(s)}+d_x^{(i)}}2 = -\frac{k_0'\Omega_-q_{x-}}{k_0^2}L
\end{equation}
and shows the influence of dispersion on the spatial walk-off in the $x$ direction. A similar quantity $y_\mathrm{wo}$ is obtained from Eq.~(\ref{xwo}) by replacing $q_{x-}$ with $q_{y-}$. Finally,
\begin{equation}\label{rho}
\rho_p=\left(\frac{n_z(\omega_0)^2}{n_o(\omega_0)^2}-\frac{n_z(\omega_0)^2}{n_e(\omega_0)^2}\right)\sin{\theta_0}\cos{\theta_0}
\end{equation}
is the pump walk-off angle in the $y$ direction.

The function $\sinc(x)$ takes the maximal value 1 at $x=0$ and changes significantly from this value when $x$ is comparable to $\pi/2$ rad. It means, that for calculating the squeezing kernel, Eq.~(\ref{eq:Kernel+}), the linear in $(\vec{q}_+,\Omega_+)$ terms in Eq.~(\ref{eq:Delta3}) can be disregarded if they are much less than $\pi/L$. The values of $|\vec{q}_+|$ and $|\Omega_+|$ in the squeezing kernel are limited to the standard deviations of the spatio-temporal intensity distribution of the pump pulse, $q_p=1/w_p$ and $\Omega_p=\sqrt{2\ln2}/\tau_p$ respectively, where $w_p$ is the waist of the pump beam, focused at the center of the crystal, and $\tau_p$ is the duration of the pump pulse. As consequence, we obtain the following conditions of smallness of the linear terms in the phase mismatch: $w_p\gg w_\mathrm{NPMPA}$ and  $\tau_p\gg\tau_\mathrm{NPMPA}$, where
\begin{eqnarray}\label{d0}
w_\mathrm{NPMPA} &=& \frac1\pi\max\left\{|x_\mathrm{wo}|,|y_\mathrm{wo}+\rho_pL|\right\}, \\\label{tau0}
\tau_\mathrm{NPMPA} &=& \frac{\sqrt{2\ln2}}{\pi}\max\tau_\mathrm{wo},
\end{eqnarray}
with maximization over all signal components.

In a crystal with positive dispersion, like the BBO crystal considered in Sec.~\ref{sec:sim}, the group velocity of the subharmonic is higher than that of the pump, i.e. $k_0'<k_p'$. It means that the maximal walk-off time is reached at degeneracy, $\max\tau_\mathrm{wo} = |k_0'-k_p'|L$. For the configuration  of Sec.~\ref{sec:sim} we obtain $\tau_\mathrm{NPMPA}=141$ fs.

The maximal values of $|x_\mathrm{wo}|$ and $|y_\mathrm{wo}|$ are reached at the components with the maximal spectral and spatial deflections and are determined by the size of the mirrors selecting the signal and idler beams from the cone of the downconverted light: $\max|x_\mathrm{wo}|=k_0'\Omega_\mathrm{max}q_{x,\mathrm{max}}L/{k_0^2}$ and a similar expression for $\max|y_\mathrm{wo}|$. For the configuration  of Sec.~\ref{sec:sim} we obtain $\max|x_\mathrm{wo}|=15$~$\mu$m and $\max|y_\mathrm{wo}|=0.24$~$\mu$m. The vertical walk-off distance of the extraordinary pump wave is $\rho_pL=-67$~$\mu$m and proves to be the main limitation for the considered configuration. Thus, we estimate $w_\mathrm{NPMPA}=21.4$~$\mu$m.

\section{Singular value decomposition in three dimensions \label{appendix:Mehler}}
The singular value decomposition of a double-Gaussian kernel reads \cite{Horoshko19}:
\begin{eqnarray}\label{eq:Mehler}
\frac1{\sqrt{\pi}}&&e^{-\frac14\frac{1+\xi}{1-\xi}\left(x+y\right)^2 -\frac14\frac{1-\xi}{1+\xi}\left(x-y\right)^2}\\\nonumber
&&= \sqrt{1-\xi^2}\sum_{n=0}^\infty (-1)^n\xi^n h_n(x) h_n(y),
\end{eqnarray}
where $0<\xi<1$. Equation (\ref{eq:Mehler}) follows from the Mehler's formula for Hermite polynomials \cite{Mehler1866} multiplied by $e^{-x^2/2-y^2/2}$ from both sides. Note that when the dispersion in the $x+y$ direction is smaller than that in the $x-y$ direction, which is the case of NPMPA, the factor $(-1)^n$ appears under the sum. As a consequence, one of the singular functions, let it be the right one, is equal to $(-1)^nh_n(y)$, while the left one is $h_n(x)$. The functions $(-1)^nh_n(y)$ are orthonormal and complete on the Hilbert space, as expected for singular functions. The singular values are positive and equal to $\sqrt{1-\xi^2}\xi^n$.

The right-hand side of Eq.~(\ref{eq:J0}) is a product of three double-Gaussian kernels. Applying Eq.~(\ref{eq:Mehler}) to each of these kernels, we obtain the singular value decomposition of JSA. In particular, for the $t$ dimension, we denote the parameter by $\xi_t$, put $x=\tau\Omega$, $y=\tau\Omega'$  and obtain from comparing Eqs.~(\ref{eq:J0}) and (\ref{eq:Mehler})
\begin{eqnarray}\label{xi1}
\frac{1+\xi_t}{1-\xi_t}\tau^2 &=& \frac1{\Omega_p^2},\\\label{xi2}
\frac{1-\xi_t}{1+\xi_t}\tau^2 &=& \frac1{\mu^2\Omega_\mathrm{max}^2}.
\end{eqnarray}
Excluding $\xi_t$ from these two equations, we obtain $\tau=(\mu\Omega_\mathrm{max}\Omega_p)^{-\frac12}$, as indicated in Sec.~\ref{Gaussian}. Substituting this value of $\tau$ into Eq.~(\ref{xi1}), we obtain $\xi_t$ as in Eq.~(\ref{xi}).

Repeating this procedure for the $x$ and $y$ dimensions, we obtain a singular value decomposition of JSA in the form of Eq.~(\ref{eq:SVD}) with the singular values given by Eq.~(\ref{eq:sl}) and the left and right singular functions
\begin{eqnarray}\label{eq:cl}
\tilde c_\ell(\eta,q_y,\Omega) &=& h_i(u\eta)h_j(vq_y)h_k(\tau\Omega) \sqrt{uv\tau},\\\nonumber
\tilde d_\ell(\eta,q_y,\Omega) &=& (-1)^{i+j+k}h_i(u\eta)h_j(vq_y)h_k(\tau\Omega) \sqrt{uv\tau}.
\end{eqnarray}

Passing back to the Cartesian coordinates according to Eq.~(\ref{eq:eta}), we obtain the modal functions, Eq.~(\ref{eq:cl2}).

\section{Orthonormality and completeness}\label{appendix:ortho}
Here we show that the infinite set of modal functions $c_\ell(q_x,q_y,\Omega)$, defined by Eq.~(\ref{eq:cl2}), with a composite index $\ell=(i,j,k)$ is orthonormal and complete on the space of real three-dimensional square-integrable functions.

First, we consider the orthonormality. Calculating the integral of $c_\ell(q_x,q_y,\Omega)c_{\ell'}^*(q_x,q_y,\Omega)$, we first perform the integration over $q_x$, considering $q_y$ and $\Omega$ as parameters. By a proper change of variables and due to orthonormality of the Hermite-Gauss functions, we obtain
\begin{eqnarray}\nonumber
&&\int c_\ell(q_x,q_y,\Omega) c_{\ell'}(q_x,q_y,\Omega) dq_xdq_yd\Omega \\\nonumber
&=& v\tau\delta_{ii'}\int h_j(vq_y)h_{j'}(vq_y)h_k(\tau\Omega)h_{k'}(\tau\Omega) dq_yd\Omega \\\label{eq:orthog}
&=& \delta_{ii'}\delta_{jj'}\delta_{kk'}.
\end{eqnarray}
The same result can be obtained by calculating the above integral in curvilinear coordinates $(\eta,q_y,\Omega)$, where $\eta$ is defined by Eq.~(\ref{eq:eta}), and noting that the Jacobian corresponding to the coordinate change is
\begin{equation}
\frac{\partial(\eta,q_y,\Omega)}{\partial(q_x,q_y,\Omega)} = \left|\begin{array}{ccc}
1 & q_y/q_d & -Q_0^2\Omega/(\Omega_0^2q_d) \\
0 & 1 & 0 \\
0 & 0 & 1 \end{array} \right| = 1.
\end{equation}

Second, we consider the completeness. Calculating the sum of $c_\ell(q_x,q_y,\Omega)c_{\ell}(q_x',q_y',\Omega')$, we first perform the summation over $j$ and $k$ and due to the completeness of the Hermite-Gauss functions we obtain
\begin{eqnarray}\nonumber
&&\sum_\ell c_\ell(q_x,q_y,\Omega) c_{\ell}(q_x',q_y',\Omega') \\\nonumber
&=& u\delta(q_y-q_y')\delta(\Omega-\Omega')\sum_i h_i\left(uq_x - uq_0\right)
h_i\left(uq_x' - uq_0\right)\\\label{eq:compl}
&=& \delta(q_x-q_x')\delta(q_y-q_y')\delta(\Omega-\Omega'),
\end{eqnarray}
where we have used $q_0=q_d + Q_0^2\Omega^2/(2\Omega_0^2q_d) - q_y^2/(2q_d)$. Thus, the considered set of functions is complete.

The orthonormality and completeness of the infinite set of modal functions $d_\ell(q_x,q_y,\Omega)$, defined by Eq.~(\ref{eq:dl2}), is proven in a similar way.

\bibliography{Twins-Space-Time-v3}            %
\end{document}